\documentclass[english]{IEEEtran}
\usepackage[T1]{fontenc}
\usepackage[utf8]{inputenc}
\usepackage{xcolor}
\usepackage{float}
\usepackage{units}
\usepackage{mathrsfs}
\usepackage{amsmath}
\usepackage{amssymb}
\usepackage{graphicx}
\usepackage{hyperref}       % hyperlinks
\usepackage{url}            % simple URL typesetting
\makeatletter

\floatstyle{ruled}
\newfloat{algorithm}{tbp}{loa}
\providecommand{\algorithmname}{Algorithm}
\floatname{algorithm}{\protect\algorithmname}

\usepackage{eqnarray}
\usepackage{mathrsfs}
\usepackage{epsfig}
\usepackage[english]{babel}
\usepackage{subfigure}
\usepackage{epstopdf}
\usepackage{import}
\usepackage{color}
\usepackage{colortbl}
\usepackage{cite}
\usepackage{algorithm}

\usepackage{bbm}
\usepackage{cases}
\usepackage{array}
\usepackage[english]{babel}
\usepackage{times}
\usepackage{color}
\usepackage{amsfonts}
\usepackage{psfrag}
\usepackage{fancyhdr}
 \usepackage{algorithmic}
\allowdisplaybreaks

\usepackage{dsfont}
\usepackage{babel}

\makeatother

\makeatletter
\def\footnoterule{\kern-3\p@
	\hrule \@width 2in \kern 2.6\p@} % the \hrule is .4pt high
\makeatother
\DeclareMathOperator*{\sumsum}{\sum\sum}

\usepackage{babel}
\begin{document}
\title{Reinforcement Learning Based Vehicle-cell Association Algorithm for Highly Mobile Millimeter Wave Communication}

\author{Hamza Khan,
        Anis Elgabli,
        Sumudu~Samarakoon, \IEEEmembership{Member, IEEE},
        Mehdi Bennis, \IEEEmembership{Senior~Member, IEEE},
        and Choong Seon Hong, \IEEEmembership{Senior~Member, IEEE}%
   
    \thanks{Hamza Khan, Anis Elgabli, and Sumudu Samarakoon are with the Centre for Wireless Communications, University of Oulu, 90014 Oulu, Finland (emails: \{hamza.khan,\,anis.elgabli,\,sumudu.samarakoon\}@oulu.fi).}
     \thanks{ Mehdi~Bennis (Corresponding author) is with the Centre for Wireless Communications, University of Oulu, 90014 Oulu, Finland, and also with the Department of Computer Science and Engineering, Kyung Hee University, Seoul 17104, South Korea (e-mail: mehdi.bennis@oulu.fi).}

    \thanks{Choong Seon Hong is with the Department of Computer Science and Engineering, Kyung Hee University, Seoul 17104, South Korea (email: cshong@khu.ac.kr).}%
\vspace{-0.4cm}}

\maketitle

\begin{abstract}
Vehicle-to-everything (V2X) communication is a growing area of communication with a variety of use cases. 
This paper investigates the problem of vehicle-cell association in millimeter wave (mmWave) communication networks. 
The aim is to maximize the time average rate per vehicular user (VUE) while ensuring a target minimum rate for all VUEs with low signaling overhead. 
We first formulate the user (vehicle) association problem as a discrete non-convex optimization problem. 
Then, by leveraging tools from machine learning, specifically distributed deep reinforcement learning (DDRL) and the asynchronous actor critic algorithm (A3C), we propose a low complexity algorithm that approximates the solution of the proposed optimization problem. 
The proposed DDRL-based algorithm endows every road side unit (RSU) with a local RL agent that selects a local action based on the observed input state. 
Actions of different RSUs are forwarded to a central entity, that computes a global reward which is then fed back to RSUs. 
It is shown that each independently trained RL performs the vehicle-RSU association action with low control overhead and less computational complexity compared to running an online complex algorithm to solve the non-convex optimization problem. 
Finally, simulation results show that the proposed solution achieves up to 15\% gains in terms of sum rate and 20\% reduction in VUE outages compared to several baseline designs.
\end{abstract}


\section{Introduction}

The automotive industry is experiencing a technological revolution enabled by vehicle-to-everything (V2X) communication. On the one hand, V2X communication enhances safety and efficiency of transportation by extending drivers' field-of-view (FoV) \cite{sensors}. On the other hand, emerging applications i.e., platooning, autonomous driving, collision avoidance, and dynamic map sharing are necessary for spearheading the vision of intelligent transportation systems \cite{car}. 
In this regard, a tremendous growth of traffic in vehicular communication systems ought to be handled by the next generation of mobile services (5G).
The state-of-the-art communication technology for enabling vehicular communication is dedicated short range communication (DSRC) based on IEEE 802.11p protocol. 
DSRC offers connectivity up to a range of $1000\,$m and the maximum achievable rate of $27$ Mbps \cite{dsrc_speed}. 
Field testing of DSRC with omni-directional antennas in a real system provided the maximum data rate of $6$ Mbps \cite{DSRC}. 
Researches on the other hand have advocated cellular solution for vehicular communication mainly because the infrastructure is already deployed. 3GPP has standardized long term evolution vehicular (LTE-V) as an extension of already existing fourth generation (4G) LTE standard. LTE-V integrates \emph{PC5 interface} to enable vehicle-to-vehicle (V2V) communication and the \emph{Uu interface} enables vehicle-to-infrastructure (V2I) communication. Achieving higher data rates with the currently deployed cellular standard is a challenging task since the maximum data rate of 4G system is limited to $100$ Mbps for highly mobile scenarios. Therefore, it is highly unlikely that the current technologies will fulfill the stringent requirements posed by the next generation of vehicular devices which will generate terabytes of data per hour. 5G is intended to support a diverse range of services, however, the initial solutions are expected to provide support for high throughput and low latency use cases \cite{qualcomm}. The increasing popularity of mission critical use cases are forcing network operators and service providers to ensure a high quality of experience/service (QoE/QoS) level.

The expected growth of traffic demands including V2X communication systems calls for higher spectral efficiencies, in which communication over millimeter wave (mmWave) bands has become a pivotal research interest in 5G technology.
However, the sensitivity of mmWave signals to blockages, higher pathlosses, frequent handovers in dense networks, highly mobile scenarios, and the huge difference in signal-to-interference plus noise ratio (SINR) between the line-of-sight (LOS) and non-LOS links are the major deployment challenges in mmWave communication \cite{challenges}. 
To overcome the aforementioned challenges, researchers have focused on developing efficient beamforming strategies \cite{beam}-\cite{beam1}, joint user association and radio resource allocation methods \cite{user-resources}, and enabling multi-connectivity \cite{multic}-\cite{multic1}. 

\subsection{Related Work}
Vehicle-cell association refers to the association of a vehicle with the road side unit (RSU) and it affects the system performance.  
In the currently deployed LTE/LTE-Advanced networks, the state-of-art association policy is the \emph{\textbf{maximum received power association}} \cite{assoc}, where a vehicle connects to the RSU with the maximum received power. 
The association based on the maximum received power is feasible for homogeneous networks and is not deemed as a practical solution considering the inherent nature and challenges of 5G technology. 
Emerging heterogeneous networks (HetNets) with variable cell size i.e., macro, pico, and femto requires a different approach than the traditional received power association. 
To ensure the loads are evenly balanced among the different cells, a \emph{\textbf{load balancing}} strategy is devised in \cite{load-balance} to increase the system capacity and to avoid congestion at the macro basestation due to the highest transmit power. 
An active load balancing strategy for mobile users is presented in \cite{user_7}, where the next associating basestation is predicted based on the past associations and the resources are reserved beforehand.
A similar approach for predicting the next user-cell association is presented in \cite{user_8}, where the association decision is based on a combined metric of the received signal strength, the delay, and the handover cost using nonlinear regression modeling. 
Another approach for load balancing is biasing transmit power towards increasing/decreasing cell coverage known as cell breathing \cite{breath}.
Furthermore, \emph{\textbf{energy efficiency}} is a key parameter in the design of 5G cellular networks and researchers have studied several approaches with different scenarios, assumptions, and network statistics.
A user association strategy for preserving energy in sleeping cells is presented in \cite{user_5}, where the user is associated to an active basestation on the basis of maximum mean channel access probability. 
Authors in \cite{user_1} study an energy efficient user association problem in heterogeneous networks with concave utility function. 
Authors in \cite{user_4} investigate the design of a user-cell association policy with the aim of maximizing the weighted sum of energy efficiency for multiple-input multiple-output (MIMO) enabled cellular networks. 
The joint design of energy efficient framework for optimizing the user association and resource allocation under network stability constraints is studied in \cite{user_6}. 
Furthermore, the works in the literature investigate \emph{\textbf{optimizing network utilities}} based on different optimization objectives such as minimizing the transmit power, minimizing a weighted combination of transmit power and resource allocation, and the joint design of user association and transmit power optimization. 
A joint transmit power and resource allocation approach enabling ultra-reliable low-latency
communication (URLLC) in vehicular networks is proposed in \cite{Sumudu_FL}.
The user-cell association problem is formulated as a convex utility function and solved in a centralized manner using the sub-gradient algorithm in \cite{user_2}. 
The joint design of user association and transmit power selection for massive MIMO heterogeneous networks under imperfect channel state information (CSI) is studied in \cite{user_3}. 
The user-cell association problem for two-tier networks operating at high frequency and mmWave frequency is studied in \cite{user_9}. 
Authors in \cite{user_10} study the performance of online algorithms for the multi-tier multi-cell user association problem. 
The URLLC use case is studied from the perspective of a scalable framework, which takes into account delay, reliability, packet size, network architecture, and topology in \cite{ Mehdi_tail_risk}.
However, associating a mmWave user with a basestation without considering the mobility may result in frequent handovers hereby, degrading the overall system performance. 
Recently, the \emph{\textbf{deep learning}} paradigm has gained interest of researchers, followed by exciting results in fields i.e., vehicular communication, user-cell association and communication at mmWave.
Edge computing is an emerging concept and authors in \cite{elbamby2019wireless} discuss the applications that must be provided at the network edge with special emphasis on URLLC enabled services i.e, V2X communication. 
A knowledge-driven edge computing mechanism using the asynchronous actor critic (A3C) based learning algorithm for vehicular networks is studied in \cite{kd}. 
A centralized reinforcement learning based resource allocation solution for out-of-coverage vehicular users is proposed in \cite{Rl-scheduler}.
Authors in \cite{cell_association_2} have used recurrent neural network to predict the next associating basestation with which a mobile node will connect. 
A deep learning based user-cell association problem for sum-rate maximization is considered in \cite{cell_association_1}, where the input to the neural network (NN) is only the geographical position of the users. 
The authors in \cite{cell_association_1} show that the association problem can be solved in a computation efficient manner using deep learning framework, where the optimization based solution has computational complexity of $KM^2$ per basestation ($K$ is the users and $M$ is the number of basestations) compared to $2KM$ computations of artificial NN.
Furthermore, the author in \cite{v2x-a3c} proposed a vehicle position control solution using the A3C framework to solve the LOS blockage problem in mmWave vehicle-to-vehicle (V2V) relaying.
The aforementioned works investigate the user-cell association problem with different objectives such as energy efficient user-cell association, joint design of user-cell association, resource allocation and joint design of transmit power, user-cell association. However, a common feature of all optimization-based studies is the requirement of intensive computation to perform the re-association under varying network conditions, i.e., propagation characteristics, fading and mobility. Hence, above solutions may incur huge computational and communication overhead during the re-association, specially for MIMO networks with multiple users and basestations. Therefore, the solutions in the existing literature may not always be feasible to implement under mmWave channel fluctuations due to the mobility. On the other hand, deep reinforcement learning (DRL) is a new paradigm for effective decision making and, compared to state-of-the-art optimization techniques, it can operate in a computationally efficient manner \cite{cell_association_1}. Another advantage of DRL approaches is that they can operate efficiently without relying on predefined models that characterizes the environmental statistics. 

\subsection{Our Contribution}
The main contribution of this paper is the design of a novel low complexity machine learning based vehicle-RSU association algorithm in mmWave communication networks. 
We formulate an optimization problem that maximizes the average sum rate of all the vehicles and at the same time minimizes the probability of events in which the average rate fall below a predefined threshold.
In our proposed approach, we have utilized multiple RSUs serving multiple vehicles with low signaling overhead in a distributed way. 
Due to the assumption of mobility and the use of mmWave, adopting conventional optimization techniques for vehicle-RSU association incurs huge computational overhead. 
To solve the formulated problem, we adopt the RL framework to approximate the solution of the network-wide optimization problem with low computational and signalling overhead compared to the centralized approaches. 
We have utilized the approximation ability of NN to mimic the input-output response, since NN are capable of approximating linear/non-linear functions up to desired accuracy. 
The distributed setting in our work allows each RSU to run an independent NN-based solver, which makes the decision entirely based on its own observed state. 
Furthermore in our proposed approach, the computationally intensive tasks are performed \emph{offline} and the validation tasks are performed \emph{online}. 
Here, offline training is utilized to approximate the solution of the NP-hard problem without strict time constraints using a large dataset that captures the environment dynamics. Offline training alleviates the requirement of intensive computation in the scenarios where the training time of NN is beyond the operational timescales. 
On the other hand, online training offers the capability of on-the-job learning through trial and error approach. In online training, the agent builds a model after repeated interactions over time and is limited by strict constraints of the operating environment. To account for the computation overhead in the online learning phase, the transmission time slot is divided into computation and transmission times.
Interestingly, the results yield higher value of the objective function when the NN is trained offline and yields comparable performance to the centralized optimization based approach when the RSUs are deployed without offline training.
Simulation results show that the proposed solution reduces the outages in terms of VUE rates dropping below a predefined threshold by $20$\% and increases the sum rate by $15$\% compared to several state-of-the-art baseline models. 
An exemplary application for this scenario is video streaming, where the users are watching different videos with strict delay constraints. 
In this case, the user's QoE is proportional to the achieved rate but in concave manner \cite{qoe_concave}. 
Therefore, the first priority is to guarantee the minimum threshold rate per user, and afterwards improving their individual rates.

The rest of the paper is structured as follows. Section II explains the system model and presents the problem formulation. The RL based vehicle association scheme is discussed in Section III. Simulation setting and the performance analysis are presented in Section IV and V respectively. Finally, the concluding remarks are provided in Section VI.  

\emph{Notations}: We will use boldface lower case letter $\textbf{x}$ and boldface upper case letter $\textbf{X}$ to represent vectors and matrices respectively. The cardinality of set $\mathcal{X}$ is denoted as $X$ and $\textbf{x}^H$ denotes the Hermitian or the conjugate transpose of vector $\textbf{x}$.

\section{System Model and Problem Definition}

\begin{figure}[hbtp]
	\centering
	\includegraphics[width=0.48\textwidth]{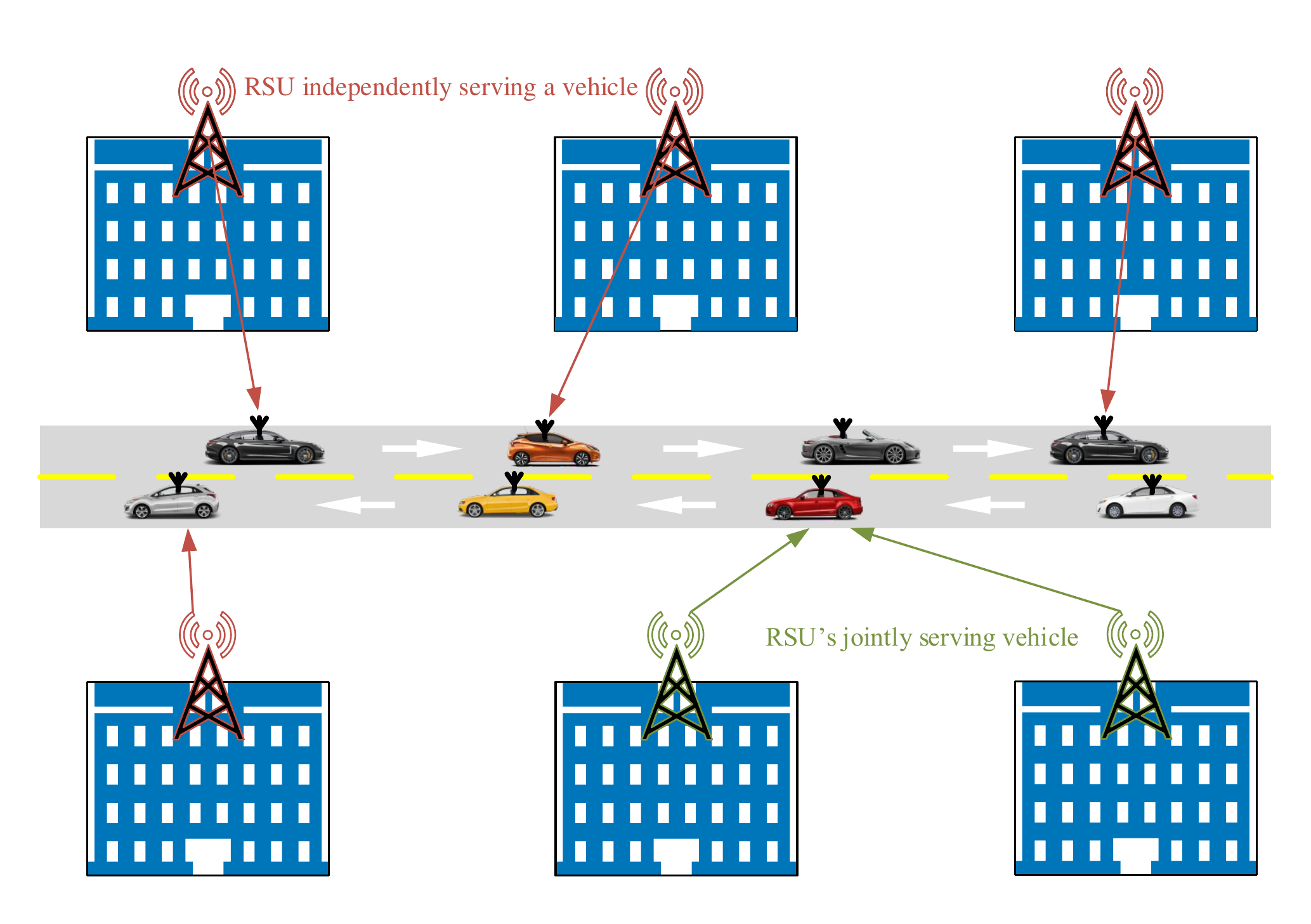}
	\caption{\label{Layout}Layout of the mmWave vehicular network.}
\end{figure}

We study the downlink system with single-input multiple-output (SIMO) transmission consisting of a set $\mathcal{B}$ of $B$ RSUs and a set $\mathcal{V}$ of $V$ vehicles. We consider a mmWave communication system, where each vehicle can be served by one or several RSUs. Illustration of the multi-RSU scenario operating in the downlink direction and situated alongside a two lane road is presented in Fig. \ref{Layout}.
Each RSU is equipped with $N_t$ transmit (Tx) antennas and each vehicle is equipped with $1$ receiver (Rx) antenna. For simplification we assume that each RSU has a single RF chain and beamforming is performed in the analog domain. In the next subsections, we explain the mmWave channel model, the segmentation of time slot and the problem formulation.

\subsection{Channel Model}
\begin{figure}[hbtp]
	\centering
	\includegraphics[trim=2 2 2 2,clip,width=0.5\textwidth]{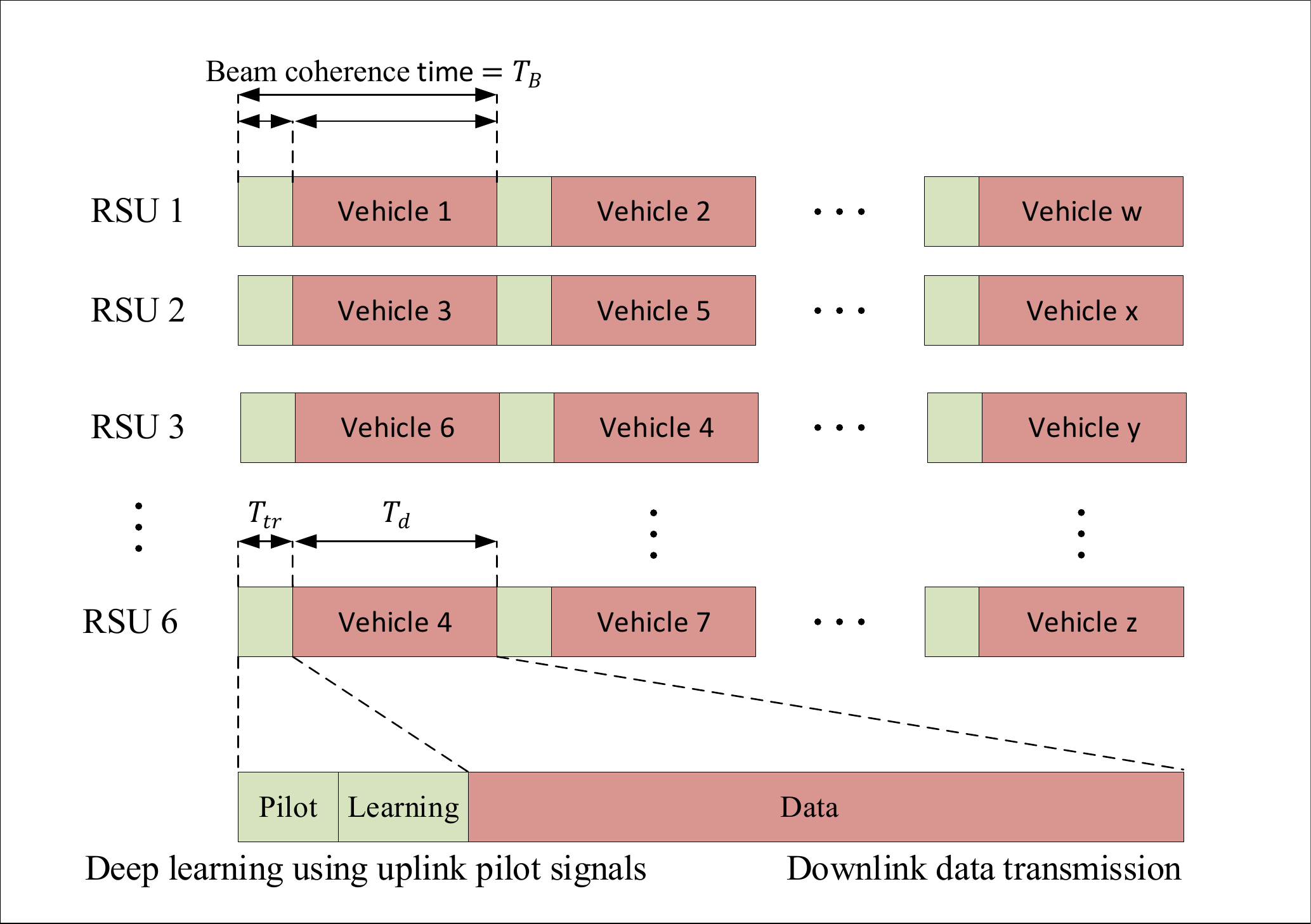}
	\caption{\label{timing}Timing diagram of computation and transmission phases.}
\end{figure}

Considering the generic mmWave channel with $L$ paths \cite{mmWave}, the frequency domain channel vector from RSU $b$ to vehicle $v$ is given as:
\begin{equation}
\label{channel}
\boldsymbol{h}_{bv} = \sqrt{\frac{N_t}{L}}\sum_{l=1}^{L} \alpha_{l} u_b(\theta_l^{\text{AoD}},\phi_l^{\text{AoD}})
u_v(\theta_l^{\text{AoA}},\phi_l^{\text{AoA}}),
\end{equation}
where $\alpha_{l}$ represents the amplitude gain of the $l^{\text{th}}$ path component. $u_b(.)$ and $u_v(.)$ represent the RSU and vehicle antenna responses at azimuth and elevation angles respectively, and $\theta_l^{\text{AoA}},\theta_l^{\text{AoD}}$ represent the angle of arrival and angle of departure of the $l^{\text{th}}$ path component. We consider a block-fading channel model, so $\boldsymbol{h}_{bv}$ is assumed to be constant over the channel coherence time $T_c$. Moreover, vehicle association with RSU at time slot $t$ is indicated by $z_{bv}(t) \in \{0,1\}$, i.e., $z_{bv}(t) = 1$ when vehicle $v \in \mathcal{V}$ is served by RSU $b \in \mathcal{B}$ at time slot $t$ and 0 otherwise. The achievable rate of vehicle $v$ at time slot $t$ is given by, 
\begin{equation}
    \label{rate}
    r_{v}(t)\ = \omega \log_2 \bigg(1 +  \frac{ \sum_b  p_{bv}  z_{bv}(t) |{f}_{bv}{\boldsymbol{h}_{bv}}|^2}{\sigma^2 + \sum_{v'} \sum_{b'} p_{b'v'}   z_{b'v'}(t) |{f}_{b'v'}{\boldsymbol{h}_{b'v}}|^2} \bigg),
\end{equation}
where $p_{bv}$ is the transmission power, $\omega$ is the bandwidth, and ${f}_{bv}$ is the analog beamformers from RSU $b$ to vehicle $v$. Moreover, $v' \in \mathcal{V} \backslash \{v\}$ denotes any other vehicle except $v$, $b' \in \mathcal{B} \backslash \{B_v\}$ represents all interfering RSUs and $\sigma^2$ represents the noise power. However, since in mmWave communications, some time slots are reserved for beam training and alignment \cite{beam_align}, the effective rate is lower than the rate specified by \eqref{rate}. Before we define the effective rate, we explain the beam training and alignment scheme we consider in this paper. 
We adopt a similar strategy to what has been proposed in \cite{facebook}, where the beam training is performed by using the pilot signals received on the uplink. The pilot signals received at different basestations are combined with the beamforming vectors and are fed to the cloud processor, which designs the coordinated beamformers to maximize the achievable rate. In \cite{facebook}, the coordinated beamformers are designed to serve a single user.
However, In contrast to that scheme, we consider that one vehicle can be served with multiple RSUs based on their locally observed states as shown in Fig. \ref{timing} i.e., the RSUs can choose to serve different or same vehicles at each time. 

\subsection{Time Slot Segmentation}
In this work, we are utilizing mmWave bands for mobile vehicles and the communication can be performed every channel coherence time or the beam coherence time, where the channel coherence time is much shorter than the beam coherence time. Therefore, we consider the communication according to the beam coherence time $T_{\text{B}}$, which is divided into two periods: (i) training and beamforming period ($T_{\text{tr}}$) and (ii) data transmission period ($T_{\text{d}}$) as it is the case in  \cite{facebook}. However, in every $T_{\text{tr}}$ period, vehicles send uplink pilot signals using omni-directional antenna in a sequential manner (one after the other), and the deep learning algorithm proposed in \cite{A3C} is used to find the best associating vehicle from each RSU $b$. The algorithm in \cite{A3C} includes the advantages of both the policy based and the value based learning techniques. The actor network in this framework formulates a policy and the critic network criticizes that policy. The detailed description of the reinforcement learning framework is explained in Section III. Moreover, considering the training and beamforming period, the effective rate $R_v(t)$ achieved at vehicle $v$ in a time slot $t$ is expressed as:
\begin{equation}
    \label{actualrate}
    R_{v}(t) = \bigg(1 - \frac{T_{\text{tr}}}{T_{\text{B}}}\bigg) r_v(t).
\end{equation}

\subsection{Problem Formulation}
The provisioning of uninterrupted service to mobile users is a challenging use case for 5G wireless networks. In this regard, the aim of this work is to enable mmWave wireless connectivity in high mobility scenario. To realize this in practice, we formulate an optimization problem with the goal of optimizing the tradeoff between maximizing the sum of average rate of all the vehicles while minimizing the probability that the average rate per vehicle falls bellow a predefined threshold. Optimizing only over the former metric may result in uneven resource allocation, yielding unacceptable low data rates for some vehicles. This is prevented by ensuring reliability in terms of maintaining the average rate per vehicle above a threshold rate. Our metric is: 
\begin{equation}
\label{objective0}
\hspace{-0.15cm} f(\mathbf{z}) =\lim_{T\rightarrow \infty} \sum_{v=1}^V\sum_{\tau=1}^t \frac{R_{v}(\tau)}{Vt} - \lambda \bigg( \textbf{Pr}\bigg\{\sum_{\tau=1}^t \frac{R_{v}(\tau)}{t} < R_v^{\text{Th}}\bigg\} \bigg), 
\end{equation}
where $\lambda$ controls the trade-off between maximizing the sum of expected rates and maintaining the probability of falling below $R_v^{\text{Th}}$ as low as possible
and $\mathbf{z(t)} = [z_{bv}(t)]_{b\in\mathcal{B}}^{v\in\mathcal{V}}$ for all $t$. Formally, the overall optimization problem is cast as:
\begin{subequations}
\label{objective}
\begin{eqnarray}
\underset{\mathbf{z(t)}}{\text{maximize}} && f(\mathbf{z(t)}),  \\
\text{subject to} && \eqref{rate} , \eqref{actualrate},  \\
 &&  \label{cons4}\sum_{v=1}^{V} z_{bv}(t)  = 1 \quad \forall b\in\mathcal{B}, \forall t,  \\
&& \label{cons5} z_{bv}(t)  \in \{0,1\} \quad \forall b\in\mathcal{B}, \forall v \in\mathcal{V}, \forall t 
\end{eqnarray}
\end{subequations}

Here, \eqref{cons4} and \eqref{cons4} imply that each RSU can at most serve one vehicle. The optimization problem defined in \eqref{objective} has integer constraints, and in general discrete optimization problems are NP hard \cite{anis}. Due to the fast dynamics in the vehicular networks, solving (5) after each beam training period has to be carried out with low latency and computation overhead. The vehicular user association techniques in the existing literature assume large coherent times and ideal cooperation among basestations while neglecting overheads, in which case these solutions become inapplicable for mmWave applications. To address this issue we resort to machine learning tools for performing the vehicle-RSU association task in a computationally efficient manner, which is discussed in details in the upcoming section.

\section{Reinforcement Learning Based Vehicle-RSU Association}

Machine learning based solutions have outperformed the start-of-art in a variety of applications. Applying machine learning tools in a mobile environment is motivated by the fact that it can learn the environment geometry using the past experience and the environment statistics \cite{facebook}. In particular, the RL-based solution are preferred over the conventional optimization techniques due to the reduction in associated overhead making it a prominent solution to support highly mobile scenarios \cite{cell_association_1}. 
A major challenge in using optimization techniques to solve distributed mobile wireless network problems in which the environment is highly variable is the computation complexity (the high running time). Such techniques rely on information exchange and require running high complexity algorithms yielding a tradeoff between overhead and performance. On the other hand, RL can convert the problem to state-action mapping in which given a certain state as an input, RL agent performs an action based on offline training. RL based algorithms do not require any model before hand, instead they learn the model through interactions with the environment. RL has produced promising results in various fields i.e., power control, edge computing and caching \cite{xianfu}. In this paper, we consider using distributed deep reinforcement learning (DDRL) to solve the problem of vehicle-cell association in highly mobile environment. In this section we explain the Markov decision process (MDP) formulation for our proposed problem and we present our DDRL-based algorithm that we propose to approximate the solution of the proposed integer programming problem.

\subsection{MDP Formulation}
Reinforcement learning follows the similar idea as MDP, where the applied decision have affect on the partially random outcome. MDP is used to express the environment for RL problems and here we explain the relation of the MDP with the assumption of mobility and the use of mmWave in vehicular networks.
Let us define the channel observed by RSU $b$ at time instant $t$ for all the vehicles in the network as $\boldsymbol{h}_b^t = (h_{bv}^t ; v \in \mathcal{V})$, the experienced rate of vehicle observed by each RSU is $\boldsymbol{R}^t = (R_{v}(t) ; v \in \mathcal{V})$, the probability of threshold violation of each vehicle can be defined as $\zeta_v(t) = \text{Pr}\{\frac{1}{T}\sum_{\tau=1}^t R_v(\tau)< R_v^{th}\}$ and the violation probability observed by each RSU is $\boldsymbol{\zeta}^t = (\zeta_{v}(t) ; v \in \mathcal{V})$. 
At each time instant, the local state observed by each RSU $b \in B$ can be described by $\boldsymbol{s}_b^t = (\boldsymbol{h}_b^t, \boldsymbol{R}^t, \boldsymbol{\zeta}^t)$, where the state includes the observed channels, the experienced rates, and the probability of violations of all the vehicles. 
Wireless channels are often highly non-stationary and the assumptions of mmWave makes the channel variations more abrupt. 
Moreover, vehicular mobility adds more randomness in the mmWave channel making it intractable.  
This change in the channel translates into the experienced rates observed by the vehicles and the threshold violations, hereby impacting the entire state. 
The state behaviour is captured by the Morkovian property corresponding to the class of memoryless MDPs, due to the assumption of mobility, mmWave and multiple serving RSUs which make the state $\boldsymbol{s}_b^t$ entirely independent and random.

\textbf{State}: The state $\boldsymbol{s}_b^t$ of the RL agent $b$ consists of the aforementioned variables required for decision making along with the history of last $k$ channels. In summary, the state $\boldsymbol{s}_b^t$ is represented as:
\begin{description}
    \item[$\bullet$] The last $k$ channel observations $\boldsymbol{h}_b^t,...,\boldsymbol{h}_b^{t-k+1}$ 
    \item[$\bullet$] The threshold violation indicator $\boldsymbol{\zeta}^t$. 
    \item[$\bullet$] The experienced rate of vehicles $\boldsymbol{R}^t$.
\end{description}

In particular, the observed state $\boldsymbol{s}_b^t$ by each RSU $b$ is aimed to design a control policy $\boldsymbol{\pi_b} = (\pi_{z_b})$, where $\pi_{z_b}$ is the vehicle association policy.
Generally, the RSUs in the network observe the input state at the beginning of each time instant and accordingly make the action (association policy) for the vehicles i.e., $\boldsymbol{\pi}(\boldsymbol{s}^t) = (\pi_{z_b}(\boldsymbol{s}_b^t); b \in \mathcal{B}) = (\boldsymbol{z}_b^t)$, where $\boldsymbol{z}_b^t = (z_{bv}^t; v \in \mathcal{V})$.

\textbf{Action}: The action of the agent is to determine the optimal vehicle-RSU association $\boldsymbol{z}_b^t$ for RSU $b$ i.e., $z_{bv} = 1, \,v \in V$.  The action performed as per the policy $\boldsymbol{\pi}(\boldsymbol{s}_b^t)$ is represented as a probability distribution $\boldsymbol{\pi}: \boldsymbol{\pi} (\boldsymbol{s}_b^t,\boldsymbol{z}_b^t) \in (0,1)$ of action $\boldsymbol{a}_b^t$ in response to state $\boldsymbol{s}_b^t$. 
The action taken by the RL agent of RSU $b$ is as follows: 
\begin{description}
\item[$\bullet$] RSU-vehicle association i.e.,  $\boldsymbol{a}_b^t = \boldsymbol{z}_b^t, \ v \in \mathcal{V}$.  
\end{description}

The problem at hand captures the tradeoff between maximizing the experienced rate and minimizing the probability of threshold violation for each RSU $b$, which can be formally defined with a reward function.

\textbf{Reward}: The reward function capturing the tradeoff of the optimization problem \eqref{objective} in this work is defined as:
\begin{multline}
       \notag \hspace*{-0.4cm} f_b(\boldsymbol{s}_b^t,\boldsymbol{z}_b^t) = { \sum_{v=1}^V\sum_{\tau=1}^t \frac{R_{v}(\tau)}{Vt} - \lambda \sum_{v=1}^V\big(\mathbbm{1}\big\{\sum_{\tau=1}^t \frac{R_{v}(\tau)}{t} < R_v^{\text{Th}}\big\} \big)}.
\end{multline}

Each RL agent observes a state $\boldsymbol{s}_b^t$ at each time instant and chooses an action $\boldsymbol{z}_b^t$ from the set of the feasible actions according to its policy $\boldsymbol{\pi}$, where the policy is the probability distribution of actions (vehicle-RSU association) and in return the RL agent receives an accumulated global reward  $f(\boldsymbol{s}^t,\boldsymbol{\pi}(\boldsymbol{s}^t)| \boldsymbol{s}) = \sum_{b \in \mathcal{B}} f_b(\boldsymbol{s}_b^t,\boldsymbol{z}_b^t| \boldsymbol{s})$. The agent tries to maximize a long term discounted reward at each time $t$ with discount factor $\gamma \in (0,1]$ i.e. $F = \sum_{t=0}^{\infty}\sum_{j=0}^{\infty} \gamma^j f^{t+j}$.

\begin{figure}[hbtp]
	\centering
	\hspace*{-1.2cm}
	\includegraphics[width=0.6\textwidth]{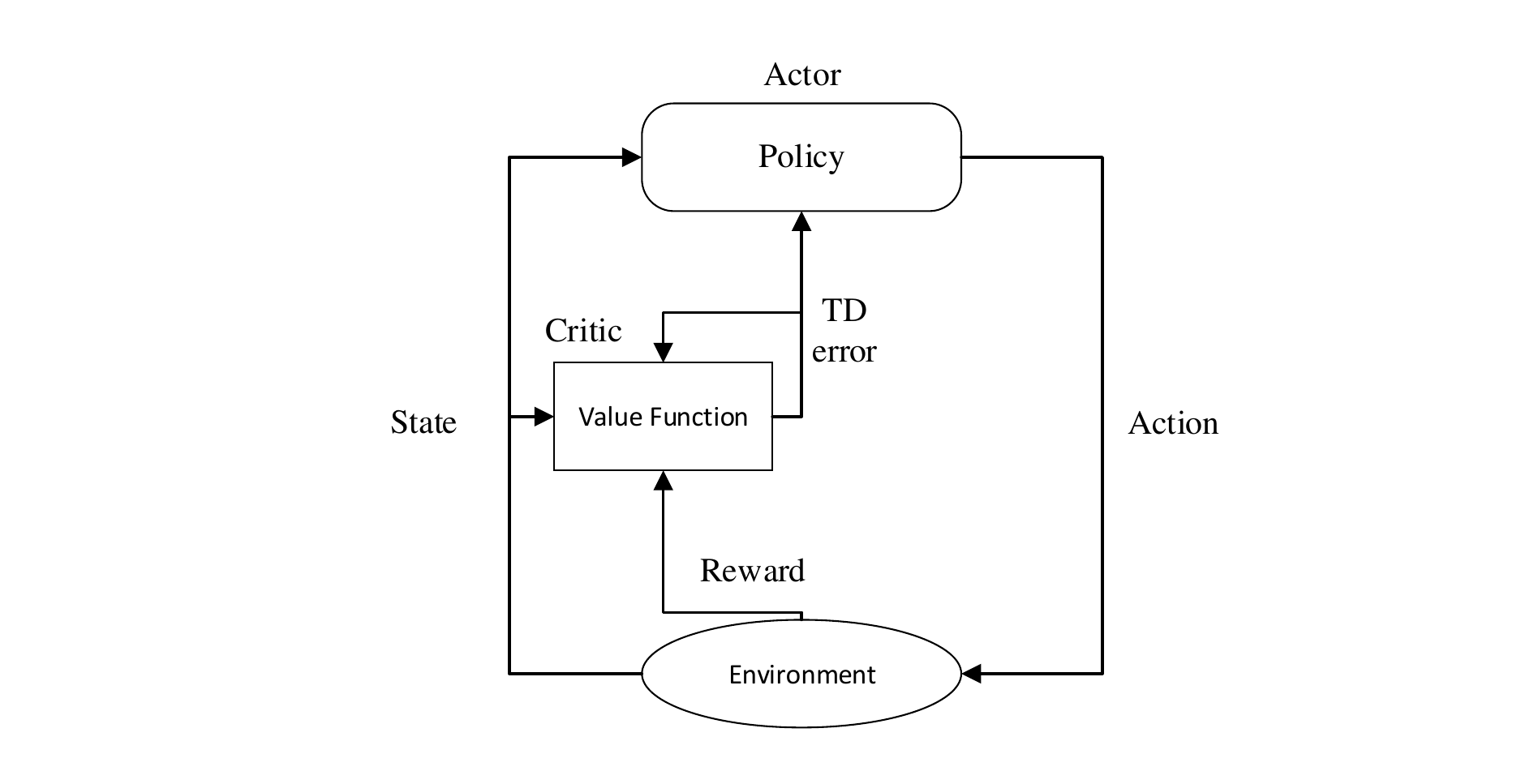}
		\caption{\label{a3c_framework}The actor-critic framework \cite{a3c_frame}}
\end{figure}

\begin{figure}[hbtp]
	\centering
	\vspace{-25pt}
	\includegraphics[width=0.5\textwidth]{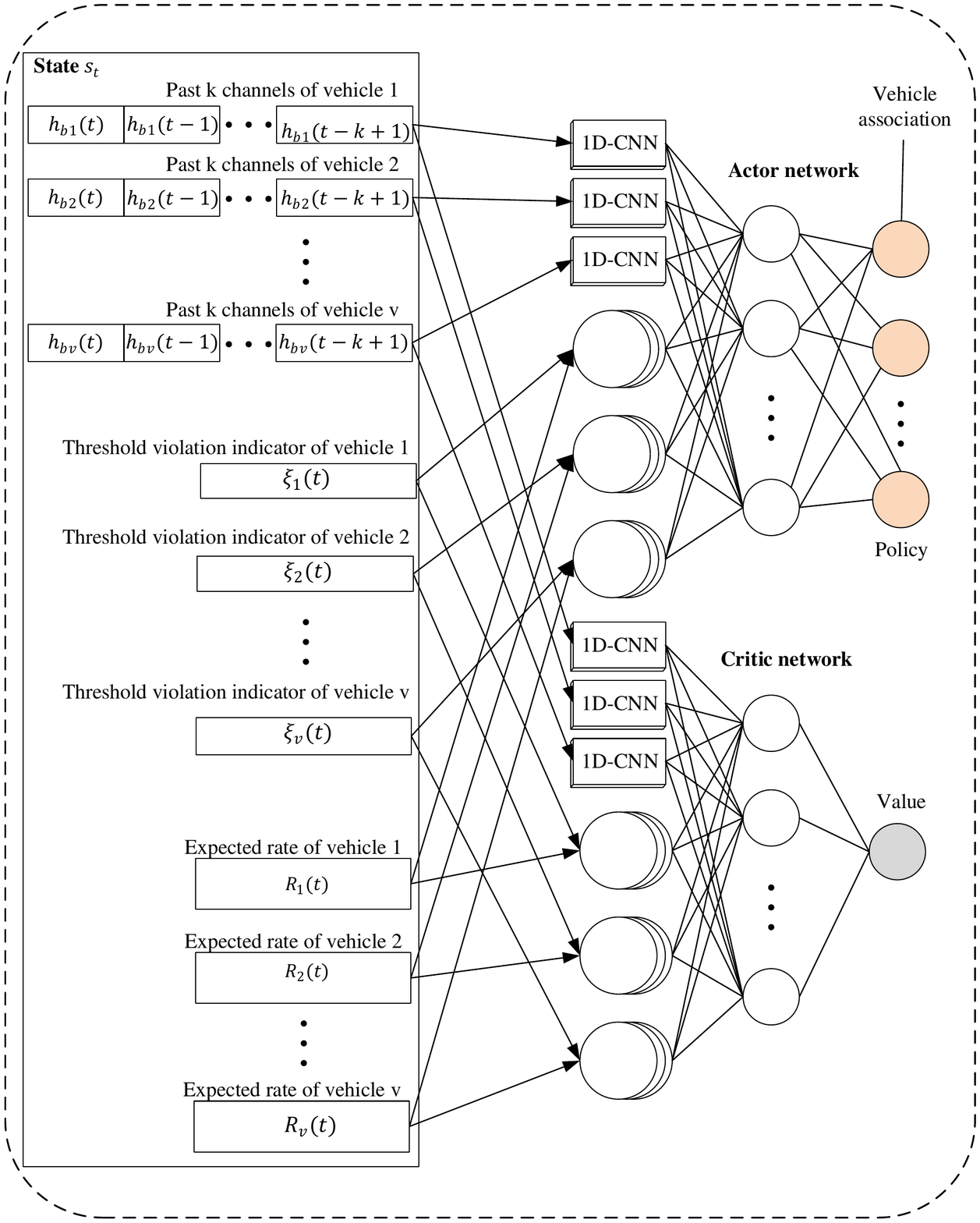}
	\vspace{-17pt}
	\caption{\label{aaac}A3C based NN architecture of RSU.}
\end{figure}

We assume that each RSU runs an RL agent, which decide the vehicle to be served at every time slot. However, every RSU forwards its action to a central entity i.e., reward aggregator, which computes a global reward and send it to all agents.  We consider the state-of-the-art A3C reinforcement learning framework \cite{A3C} in which every RL agent consists of an actor-critic pair, but with the addition of the concept of a global reward. The actor network generates a policy based on the probability distribution of the actions given the states and the critic network criticizes the policy of the actor using the temporal difference (TD) error \cite{td11}. The A3C framework utilized in this work is shown in Fig. \ref{a3c_framework} and the NN architecture of A3C framework is shown in Fig. \ref{aaac}, which is explained in more detail in the next subsection. In Fig. \ref{a3c_framework} the value function represents the critic network and the policy represents the actor network.
 
\subsection{A3C Framework}
The A3C algorithm asynchronously updates the parameter of NN using the stochastic gradient descent method. The A3C framework has two main components, \emph{actor} and \emph{critic}, which are constructed using NNs. The learning methods based on the value function are referred to as critic-only i.e., temporal difference (TD) learning. Furthermore, the actor-only learning involves the utilization of the probability distribution of the actions instead of computing the value functions.
The actor-critic framework is a combination of both the action value based and the policy based reinforcement learning, where the actor network formulates a policy based on the probability distributions of different actions and the critic network criticizes the actor policy with value function using the TD error. Since, the actor and critic network are performing different tasks this leads to a difference in the NN architecture. In case of the actor network, the output layer consists of $N$ neurons, which reflect the probability distribution of the $N$ available actions. On the other hand, the critic network has one output neuron for the value function with which it criticizes the policy of the actor network. In A3C algorithm, the policy $\boldsymbol{\pi}$ represents the actor network and the critic network estimates the advantage function, which is defined as 
\begin{equation}
    A(\boldsymbol{s}^t,\boldsymbol{a}^t) = Q(\boldsymbol{s}^t,\boldsymbol{a}^t) - V(\boldsymbol{s}^t),
\end{equation}
where $Q(\boldsymbol{s}^t,\boldsymbol{a}^t)$ is the Q-value of performing action $\boldsymbol{a}^t$ compared to the value of optimal action $V(\boldsymbol{s}^t)$ in state $\boldsymbol{s}^t$. The advantage function is a measure of how good the specific action is compared to the optimal action. The parameterization of the actor and the critic is done with their respective NNs using the parameters $\theta$ and $\theta_c$ respectively and the pseudocode for A3C algorithm is presented in Algorithm \ref{a3c_algorithm}. The parameters of the the agents are updated upon reaching the terminal state using the gradient descent method.

\floatname{algorithm}{Algorithm}
\begin{algorithm}
 \caption{A3C algorithm pseudocode}
 \begin{algorithmic}[1]
 \label{a3c_algorithm}
 \renewcommand{\algorithmicrequire}{\textbf{Input:}}
 \renewcommand{\algorithmicensure}{\textbf{Output:}}
  \STATE Initialize the parameter of learning agents $\theta$ and $\theta_c$ 
  \STATE Get initial state $\boldsymbol{s}^t$ of each agent
  \FOR {$\text{time} = 1,...,t$}
    \STATE Perform action $\boldsymbol{a}^t$ according to policy $\boldsymbol{\pi}_{\theta}(\boldsymbol{s}^t,\boldsymbol{a}^t)$.
  \STATE Forward the actions to the reward aggregator.
  \STATE Obtain global reward $f^t$ from reward aggregator.
  \STATE Observe new state $\boldsymbol{s}^{t+1}$.
  \IF{terminal state}
  \STATE Aggregate actor gradient $\theta$ w.r.t. \eqref{actor_update_theta}.
  \STATE Aggregate critic gradient $\theta_c$ w.r.t. \eqref{critic_update_theta}.
  \STATE Update parameters $\theta$ and $\theta_c$.  
  \ENDIF
  \ENDFOR
 \end{algorithmic} 
 \end{algorithm} 

The A3C algorithm works by maintaining a policy $\boldsymbol{\pi}(\boldsymbol{s}^t,\boldsymbol{a}^t;{\theta})$ along with an estimate of the value function $V(\boldsymbol{s}^t;{\theta_c})$. The policy and the value function are incremented after reaching a terminal state at which point their gradients are computed. The gradients are used to estimate the accumulated reward by observing the action trajectories of the policy. The gradient of the accumulated reward is updated as:  
\begin{equation}
\nabla_{\theta} \mathbbm{E}_{\theta}^{\pi}\big[\textstyle    \sumsum\limits_{t,j=0}^{T-1} \gamma^t f^{t+j} \big]  = \mathbbm{E}_{\theta}^{\pi} \big[\nabla_{\theta} \log \boldsymbol{\pi}(\boldsymbol{s}^t,\boldsymbol{a}^t;{\theta}) A^{\pi}(\boldsymbol{s}^t,\boldsymbol{a}^t;\theta) \big],
\end{equation}
where $A^{\pi}(\boldsymbol{s}^t,\boldsymbol{a}^t;\theta)$ is an estimate of the advantage function when following policy $\boldsymbol{\pi}$ \cite{A3C}. The advantage function  represents the difference in the expected reward obtained from following a deterministic policy compared to the expected reward obtained by following policy $\pi_{\theta}$. To compute the expected return of following a deterministic policy, we need an estimate from the critic network, which is found through exhaustive search method. However, instead of exhaustively searching for the value of deterministic policy we can update the policy parameter towards minimizing the long term discounted reward of the chosen action. In this regard, the update of policy parameter $\theta$ follows the idea of policy gradient as follows:

\begin{equation}
    \theta^{t+1} = \theta^t + \alpha \sum_{t}^{T-1} \nabla_{\theta} \log  \boldsymbol{\pi}_{\theta}(\boldsymbol{s}^t,\boldsymbol{a}^t) A^{\pi}(\boldsymbol{s}^t,\boldsymbol{a}^t),
\end{equation}
where $\alpha$ is the learning rate of the RL agent. The policy parameter update specifies the direction to increase the probability distribution $\pi$, where it effects to reinforce action that leads to more reward. To update the critic network $\theta_c$, we need to compute the value function $V_{\theta}^{\pi}(\boldsymbol{s}^t)$. The critic network in A3C based RL agents learns the estimates of the value function from the discounted reward. While, the traditional learning techniques which compute the value function become infeasible due to the huge state space. To overcome this, an alternative approach is to estimate the Q-values using NNs, which map the value of state-action pair to their corresponding Q-values. Moreover, applying NN in Q-learning scenarios may result into unstable behaviour due to correlation  in the training samples. The correlation between the training samples can be reduced by using a deep NN as a function approximator of the Q-values, where the agent explores random actions and stores the experience in a target network \cite{minh}. The experience is randomly sampled in mini-batches to break the correlation between training samples known as experience replay. However, in this work we utilize the idea of asynchronously updating the agents in a parallel manner instead of experience replay \cite{A3C}. With the help of parallelism the training samples are decorrelated into a more stationary process, as parallel agents experience different instances of the state. 
Parameter $\theta_c$ of critic network is updated according to the temporal difference method \cite{td}. 

\begin{equation}
\label{critic_update_theta}
    \theta_{c}^{t+1} = \theta_{c}^{t} - \beta \sum_{t}^{T-1} \nabla_{\theta_{c}}\big(x_t + \gamma V_{\theta_c}^{\pi}(\boldsymbol{s}^{t+1};\theta_{c}) - \gamma V_{\theta_c}^{\pi}(\boldsymbol{s}^t;\theta_{c}) \big),
\end{equation}
where $\theta_{c}^{t+1}$ is the new search direction, $\beta$ is the critic learning rate and $\gamma$ is the discount factor which represents the weights of instantaneous reward vs long term reward. The update parameter compares the prediction of the value function of current state to the value function of next state and is termed as temporal difference approach. It was found in \cite{A3C} that adding an entropy regularization term in the actor update rule encourages exploration to discover optimal policies, hereby discouraging premature convergence to suboptimal policies. The actor update rule with entropy $H$ is given as: 
\begin{multline}
    \label{actor_update_theta}
    \theta^{t+1} = \theta^t + \alpha \sum_{t}^{T-1} \nabla_{\theta} \log \boldsymbol{\pi}_{\theta}(\boldsymbol{s}^t,\boldsymbol{a}^t) A^{\pi}(\boldsymbol{s}^t,\boldsymbol{a}^t)  + \\ \eta \nabla_{\theta} H(\boldsymbol{\pi}_{\theta}(\boldsymbol{s}^t)),
\end{multline}
where $\eta$ controls the strength of entropy with high value of $\eta$ encouraging exploration and vice-versa. The RL agent in this work uses softmax output for the actor network which is the policy $\pi$ and a linear output for the critic network which is the value $V_{\theta}^{\pi}(\boldsymbol{s}^t;\theta_c)$. To train the network, the value of entropy is kept higher at the start to encourage exploring good policies and then gradually reduced over time to enforce maximizing rewards. The training methodology used in this work is explained in the next subsection.

\subsection{Training Methodology}
\begin{figure}[hbtp]
	\centering
	\includegraphics[width=0.5\textwidth]{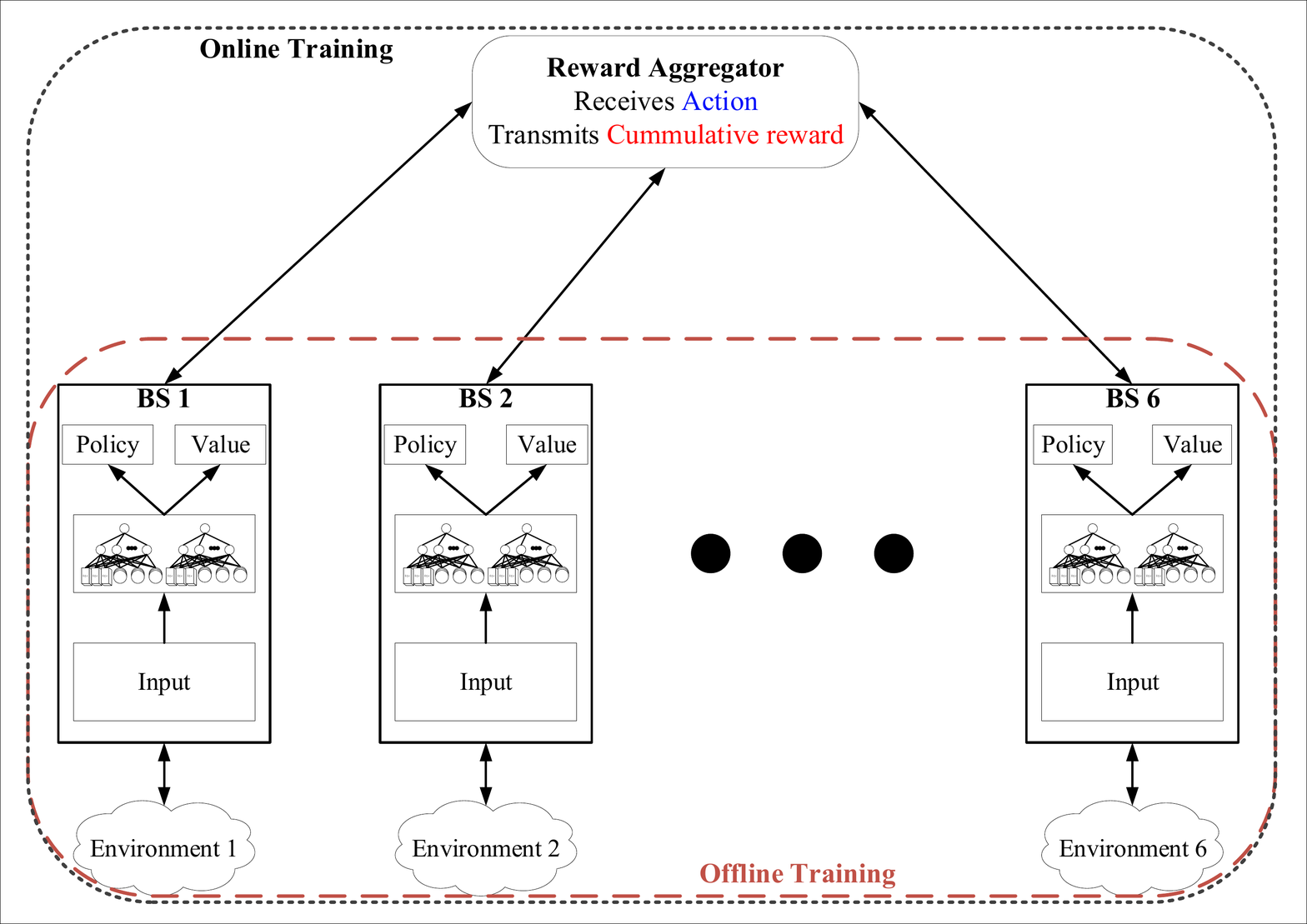}
	\caption{\label{training}Offline vs Online training.}
\end{figure}

The A3C algorithm used for training the deep NN is presented in Algorithm 1. The first step in the learning phase is the interaction with the environment i.e., observing the channels of each vehicle from RSU, the experienced rate and the threshold violations of all the vehicles as shown in Fig. \ref{training}. In this work, each RSU is modelled as an independent NN which is trained offline to reduce the overhead of 
intensive computation in learning the mapping between the input state and output actions as seen in Fig. \ref{training} and similar strategies have been followed in \cite{ml1}, \cite{ml2}. However, in the online learning phase where all the trained NNs are deployed together, the mapping of input/output is performed using a pre-trained NN without performing intensive computation \cite{ml3}. In this work, we consider a central entity referred to as \emph{reward aggregator}, which computes the global reward based on the action of all the RSUs. The accumulated global reward is propagated back to the RSUs by the reward aggregator. 

The association of vehicle with RSU is performed as per the policy of the RL agent. After each action the RL agent receives a global reward that reflects how good is the chosen action.  
In the offline learning phase, the RL agent explores different policies for which entropy weight plays an important role.  
The offline learning phase is carried out to avoid the complexity of online training and to learn the mapping of state to action (probability distribution) in a mobile vehicular environment i.e., the NN model can map the expected rates from the observed channels and can choose the best vehicle-RSU association yielding maximum reward. Once the offline learning phase has converged to an optimal policy, we deploy all the individually trained NNs in a distributed setting. The reward aggregator plays an important role due to the distributed setting of the NNs, since there is no information exchange between the RSUs and the global reward is a function of actions of all the RSUs. In the online learning phase association of each vehicle can be with a single RSU (disjoint association) or with multiple RSUs (joint association), which depends upon the individually learned policy during the offline training. In the next section we explain the simulation environment. 

\section{Simulation Environment}

For an accurate performance evaluation, it is essential to have a realistic simulation environment. 
In this regard, the following subsections discuss the simulation environment used in this work in details.

\subsection{Channel Generation}

For mmWave channel generation, we use NYUSIM channel simulator that is developed on the real word channel measurements \cite{nyusim}.
Therein, a 28\,GHz mmWave band with radio frequency bandwidth of 800\,MHz in urban micro environment and LOS connectivity is considered. 
NYUSIM generates the temporal and spatial channel impulse response (CIR) from the omni-directional channel models of NYU Wirless that uses statistical spatial channel model (SSCM) \cite{nyu_wireless}. 
It utilizes time clusters (multipath components travelling closely in the time domain and arrive at the receiver within a short excess delay window) and spatial lobes to model the omnidirectional CIR and the AoAs, AoDs power spectrums. 
Multipath components that arrive from different AoAs within a 25\,ns window are grouped in one time cluster. 
Spatial lobes on the other hand represent the direction of maximum energy. 
The real world measurement of NYUSIM indicates that time cluster can have a value between 1-6, and the spatial lobes have an average value of 2 and a maximum value of 5. 
The simulation parameters used for channel generation are listed in Table \ref{sim_param}.

\begin{table}[!tp]
\caption{Simulator parameters \cite{nyu_wireless}.}
\centering
\label{sim_param}
\begin{tabular}{l c}
\hline
\textbf{Parameter}   & \textbf{Value}             \\ \hline \hline
Scenario             & Urban micro                \\ 
Tx Power             & 30\,dBm                     \\ 
Barometric Pressure  & 1013.25\,mbar               \\ 
Humidity             & 50\%                      \\ 
Temperature          & 20$^\circ$\,C         \\ 
Polarization         & Co-Pol                     \\ 
Path Loss Exponent   & 2                          \\
\hline \multicolumn{2}{c}{ \textbf{Tx Antenna} } \\ \hline
Type      & Uniform linear array (ULA) \\ 
Elements  & 128                        \\ 
Spacing   & 0.5 Wavelength             \\ 
Azimuth   & 10$^\circ$                 \\ 
Elevation & 10$^\circ$                 \\ 
\hline \multicolumn{2}{c}{ \textbf{Rx Antenna} } \\ \hline
Type      & ULA                        \\ 
Elements  & 1                          \\ 
Azimuth   & 360$^\circ$                \\ 
Elevation & 180$^\circ$                \\ \hline
\end{tabular}
\end{table}

Using the aforementioned parameters, we have generated the channel impulse responses of mobile vehicles for different RSUs as shown in Fig. \ref{Layout}. 
During the simulation time, 2000 CIRs are generated for each VUE Rx.

\subsection{VUE-RSU Setup}

\begin{figure}[!tp]
	\centering
	\includegraphics[width=0.5\textwidth]{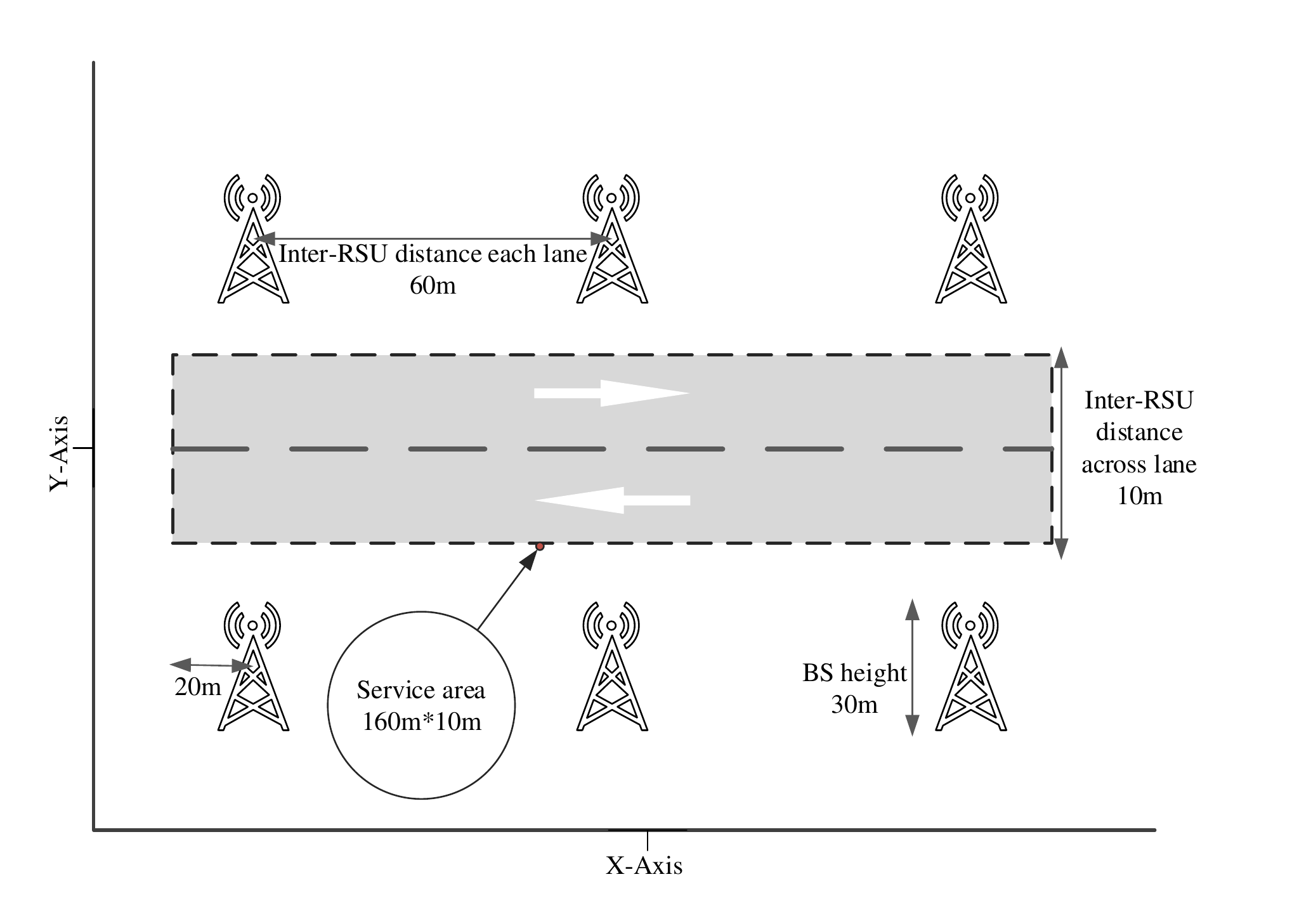}
	\caption{\label{env} Geometry of the simulation environment.}
\end{figure}

The service area consists of a road segment with the width of 10\,m and the length of 160\,m as shown in Fig. \ref{env}.
Here, 6 RSUs are installed on building rooftops at a height of 30\,m, apart from a distance of 60\,m from one another on each side of the road.
Each RSU is equipped with ULA antennas facing the street. 
The antenna array has 128 antenna elements, which use a total of 30dBm transmit power. 
The RSUs operate independently from one another without any coordination in a distributed manner. 

Each VUE is equipped with a single omni-directional antenna that is placed at a height of 1.5\,m above the ground.  
In this work, we have assumed the number of VUEs in the service area to be eight, i.e., four vehicles in each lane. 
The start point of each vehicle is chosen randomly and then the VUEs move in the direction corresponding to their lanes with an average speed of 25\,km/h. 

\subsection{Neural Network Implementation}
The state-of-the-art A3C architecture illustrated in Fig. \ref{aaac} is used in this work to devise the VUE-RSU association policy. 
Here, two NNs, the actor and the critic, deployed in an RSU use a mix of convolution layers and fully connected layers \cite{penseive}. 
In the actor network, the channels from the RSU to all vehicles are passed to a 1D convolution layer, which maintains a history of last $k$ channels. 
The indicator of  rate outages and expected rates of all VUEs are fed into the fully connected layers. 
All the layers use rectified linear activation units (ReLU). The result from input layer is aggregated at a fully connected hidden layer that is composed of 64 neurons and the aggregated data is passed on to the output layer, which applies a softmax function.
Similarly, the critic network uses the same architecture and share the same inputs, but instead of returning a softmax output it returns a value function using linear neuron as shown in Fig. \ref{aaac}. 
The training of the actor and critic network uses RMSprop optimizer \cite{rmsprop}.
To account for immediate reward vs delayed reward, a discount factor of 0.99 is used, which indicates that the current reward is influenced by 100 future iterations. 
The learning rate of the NN plays an important role in the convergence to the optimal policy. 
Learning rate specifies the magnitude of step taken towards the optimal solution i.e., small value of learning rate results slow convergence while a large value of learning rate may produce oscillations and thus, no convergence. 
In this regard, the learning rates of the actor and critic network is fixed at $0.0001$ and $0.001$. 
All these hyperparameters remain fixed in the offline and online phase. 
Although optimizing these hyperparameters based on simulation environment can lead to policies with higher rewards, such investigations are not included within the scope of this paper. 

\section{Performance Analysis}
In this section, first we analyze the performance of the proposed DRL-based VUE-RSU association solution compared to several baselines designs. Next, we study the impact of NN parameters on the performance of the proposed technique, where we have found that utilizing more training episodes results into higher reward and lower threshold violations. Moreover, the effect of hidden layers in the NN is also analyzed therin.

\subsection{{Performance of Proposed Solution}}
In this work, we consider two different schemes with the proposed design: \emph{DRL with offline training} over 2000 CIRs and \emph{DRL without offline training}.
Then, the proposed solutions are compared with three baselines:
(i) \emph{Optimization based solution}: a central controller collects the VUE-RSU channel information over the network and solves \eqref{objective} for each time $t$,
(ii) \emph{Max RSSI}: VUEs report the receive signal strength indicator (RSSI) to RSUs in which each RSU associates with the VUE with maximum RSSI \cite{assoc}
and
(iii) \emph{Proportional Fair}: each RSU observes average rates over all VUEs and make association decision based on proportional fairness criteria \cite{pf} .
In all above scenarios, 0.5\,Gbps of the target minimum rate is considered.
\begin{figure}[hbtp]
	\centering
	\includegraphics[width=0.5\textwidth]{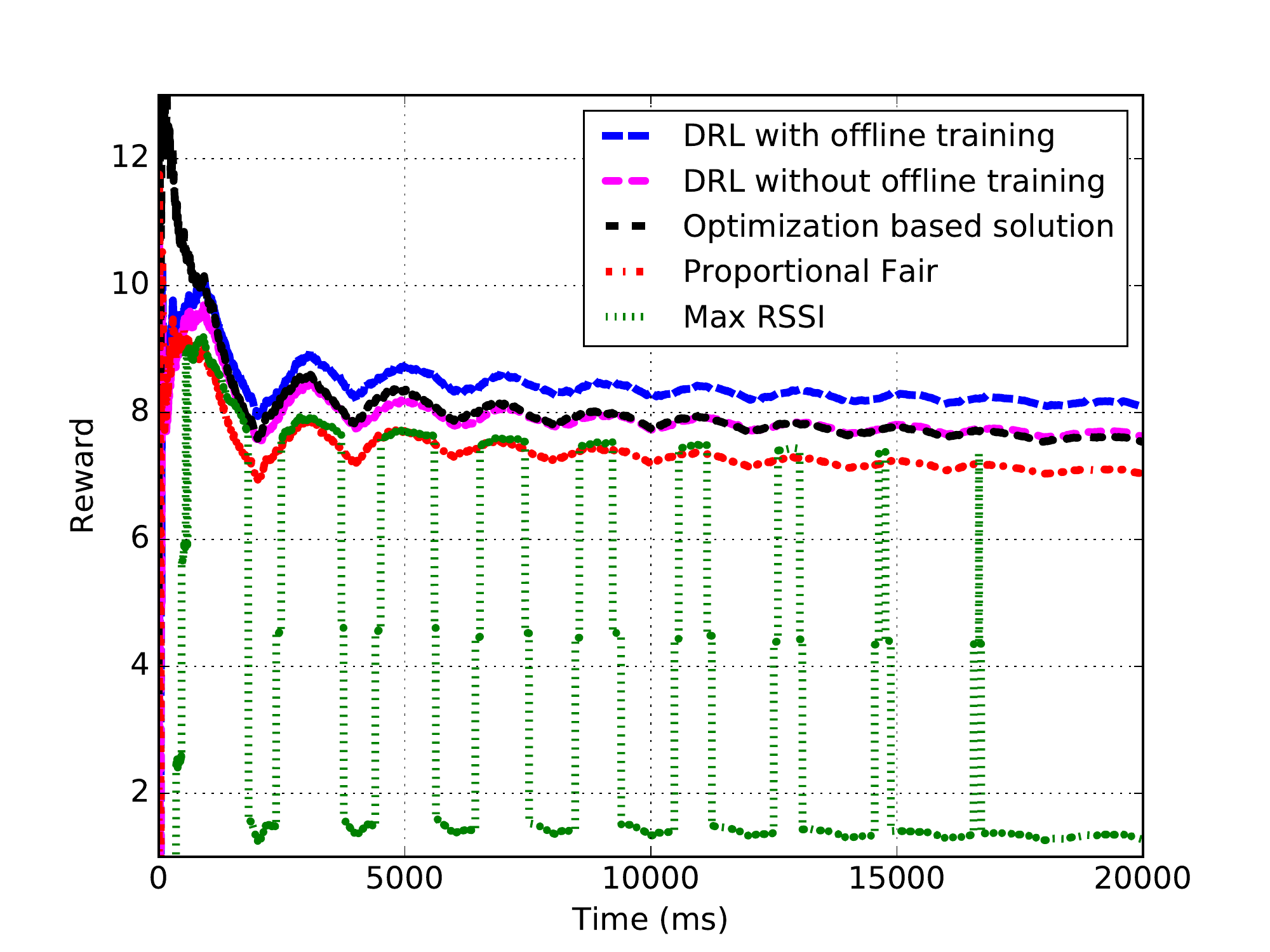}
	\caption{\label{rew}Experienced reward over time.}
\end{figure}

Fig. \ref{rew} shows the observed reward over time for the proposed RL-based solution and the baselines. The observed reward for the proportional fair method remains almost the same over time, since the goal here is to maintain an average rate for all the vehicles by following a proportional fairness policy. In this regard, the RSU pushes the average rate of all the vehicles above the threshold without maximizing the individual vehicle rate. The observed reward for the max RSSI baseline fluctuates over time, due to the fact that it opportunistically maximizes the rate whenever possible without considering the threshold violations. The myopic optimization approach achieves higher reward than the rest of the baselines and achieves the same performance as the DRL agent without offline training. The observed reward with the proposed deep learning actor critic solution with offline training outperforms all the baselines, while the reward for the DRL approach without offline training is the same as the proposed optimization based solution. 

\begin{figure}[hbtp]
	\centering
	\includegraphics[width=0.5\textwidth]{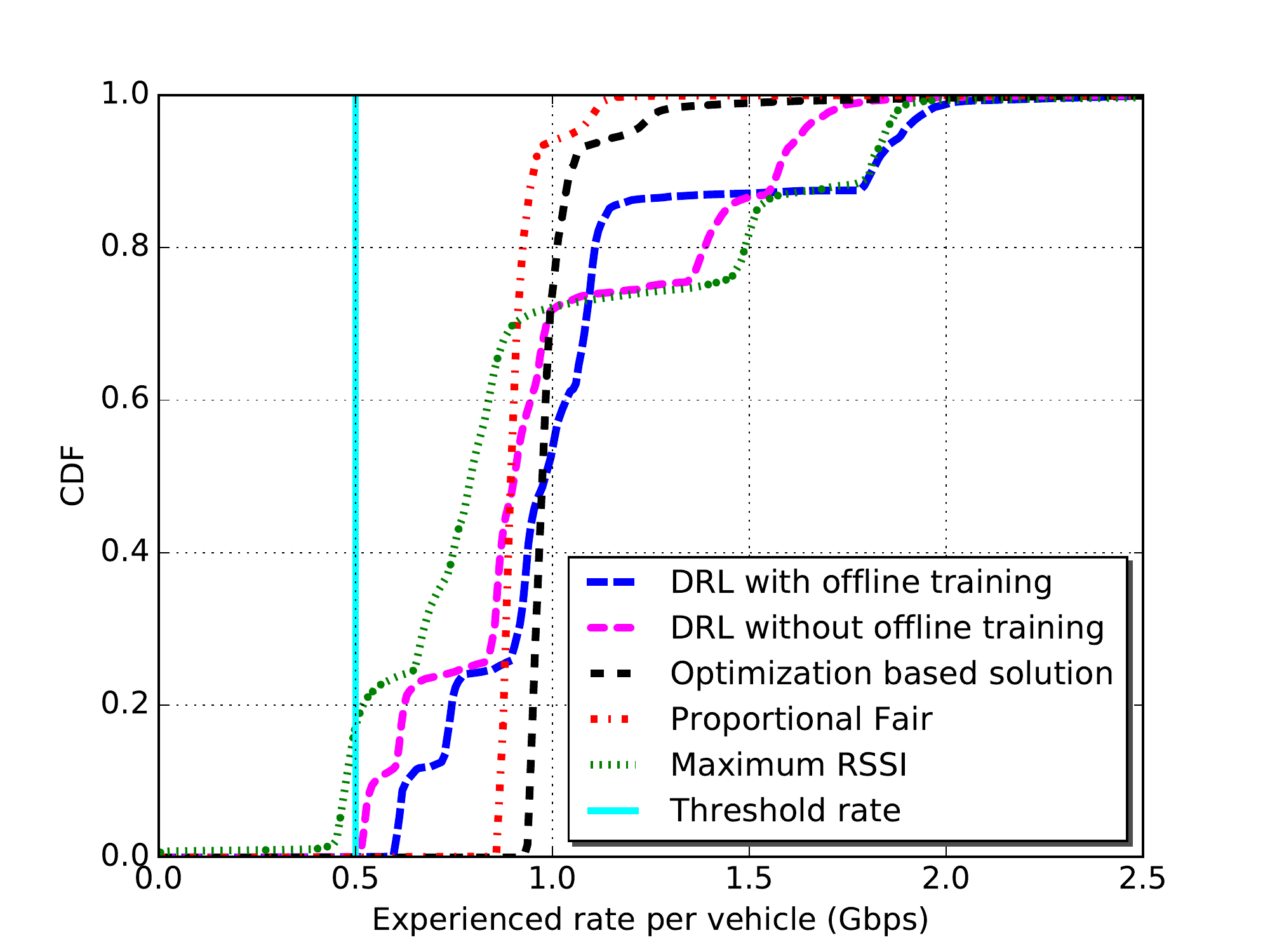}
	\caption{\label{results}CDF of the experienced rate of vehicles.}
\end{figure}

Fig. \ref{results} compares the performance of the proposed deep learning solution with the baselines in terms of the CDF of the experienced rate. Proportional fairness maintains an average experienced rate of 1.15 Gbps for all the vehicles at all times while keeping the violations at its minimum. The max RSSI performs the worst in terms of achieving a threshold rate, while maximizing the individual rates. The experienced rate for max RSSI ranges from 0.1 Gbps to 2 Gbps, with the mean value of experienced rate at 1.17 Gbps. The violation in case of max RSSI is the largest, since it opportunistically maximizes the individual vehicle rate. Moreover, the proposed A3C learning based solution with offline training achieves the highest reward as seen from Fig. \ref{rew} and this translates to the gain in experienced rate of all the vehicles as shown in Fig. \ref{results}. The expected rates of the proposed solution with and without offline training varies from 0.5 Gbps to 2.5 Gbps, with an average value of 1.35 Gbps for the former and 1.28 Gbps for the latter scenario. Moreover, the threshold criteria of 0.5 Gbps for all the vehicles is satisfied by the proposed learning approach at all times compared to the baseline vehicle-RSU association approach, which violates the minimum threshold criteria for 20\,\% of the cases.

\begin{figure}[hbtp]
	\centering
	\includegraphics[width=0.5\textwidth]{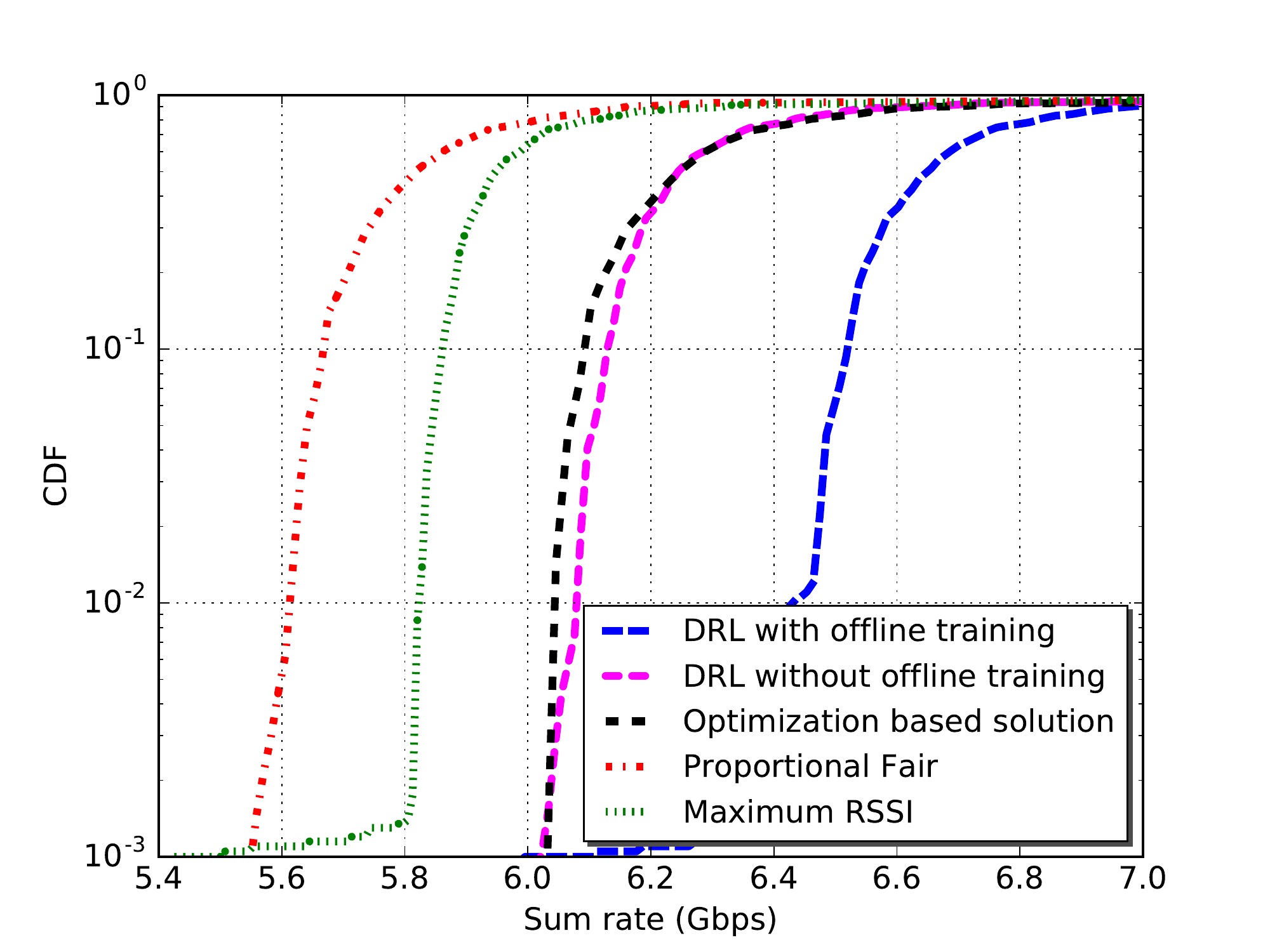}
	\caption{\label{sum_rate}CDF of sum rate of all the vehicles.}
\end{figure}

The CDF of sum rate of the all the vehicles is shown in Fig. \ref{sum_rate}. The sum rate of the proposed solution is better compared to the baselines. The sum rate of 7 Gbps is achieved for the proposed DRL with offline training approach which confirms that the proposed approach maintains an average rate of 1 Gbps as shown in Fig. \ref{results}. The performance of the sum rate of the proposed DRL without offline training is the same as the optimization based solution. Maximum RSSI performs better than proportional fair approach in terms of the sum rate as it opportunistically tries to maximize the sum rate without considering the threshold violations, which is the reason of lower reward compared to proportional fair as seen in Fig. \ref{rew}. The vehicles violate the threshold for the max RSSI as seen from Fig. \ref{results}.  
It is clear from the results that the proposed DRL scheme learns the environment geometry from the observed states at each time slot and the NN based vehicle-RSU association policy can outperform the traditional techniques.

\begin{figure}[hbtp]
	\centering
	\includegraphics[width=0.5\textwidth]{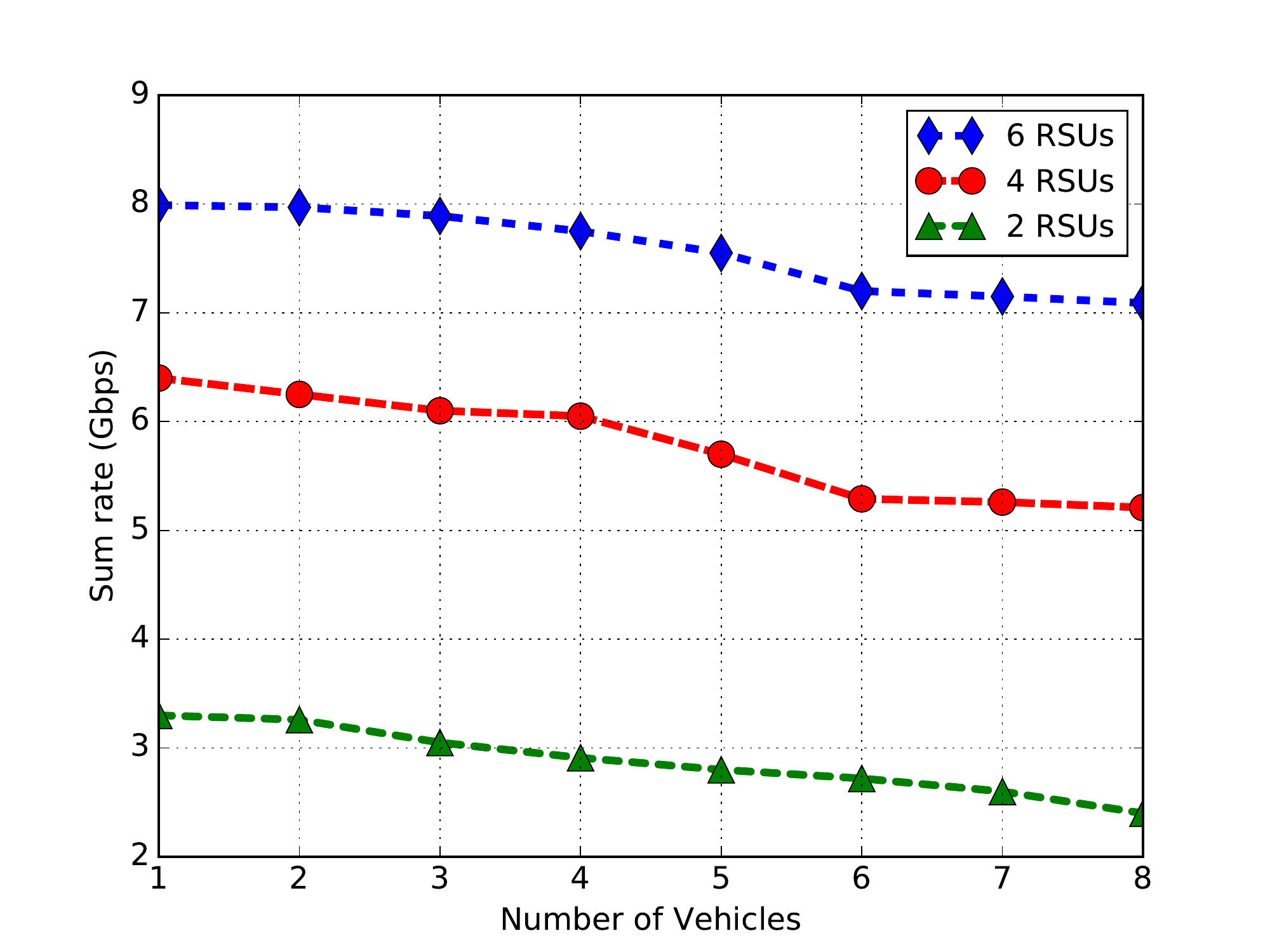}
	\caption{\label{vehsr}Effect of the number of vehicles on sum rate.}
\end{figure}

The effect of the number of vehicles on sum rate is shown in Fig. \ref{vehsr}. The sum rate decreases with the increase in the number of vehicles due to large computation times. The impact of computation time on the achieved rate is given by \eqref{actualrate}. The sum rate for 1 vehicle with 4 RSUs is 6.3 Gbps which is reduced to 5.3 Gbps, when the number of vehicles increases to 8. On the other hand, increasing the number of RSUs increases the sum rate. The sum rate for 4 vehicles with 2 RSUs is 3 Gbps, which is increased to 6 Gbps when the RSUs are doubled. Furthermore, the increase of RSUs from 4 to 6 increases the sum rate by 2 Gbps for the same number of vehicles. The decrease in sum rate with the increase in number of vehicles is associated with the overhead of computation time. 

\begin{figure}[hbtp]
	\centering
	\includegraphics[width=0.5\textwidth]{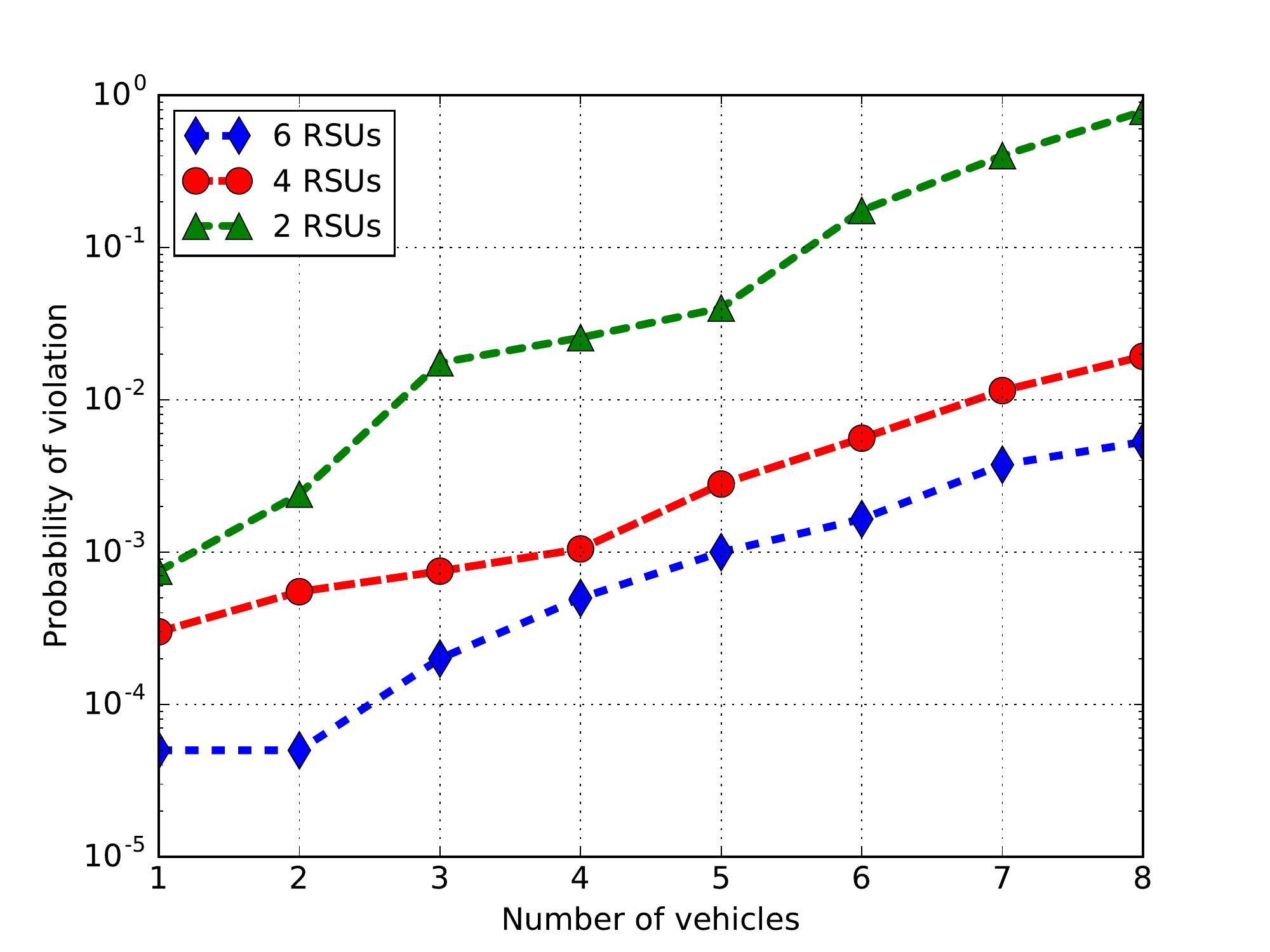}
	\caption{\label{vehvio}Effect of the number of vehicles on the probability of violations.}
\end{figure}

The objective function in this work accounts for minimizing the violations along with maximizing the sum rate. The effect on the number of violations as a function of different vehicular densities, and number of RSUs is shown in Fig. \ref{vehvio}. The number of violations increases with the increase in the number of vehicles or the decrease in the number of RSUs. The probability of violation for the scenario with 6 RSUs and 4 vehicles is $5\times 10^{-4}$, which is increased to $5.3\times 10^{-3}$ when the number of vehicles are doubled, while keeping the same number of RSUs. On the other hand, for the same number of vehicles when we reduce the number of serving RSUs we see an increase in the number of violations, which is because of the lower experienced rate as shown in Fig. \ref{vehsr}. The formulated objective function has the maximum value when there are less number of vehicles in the network and the performance gets worse with increasing vehicles. 

\begin{figure}[hbtp]
	\centering
	\includegraphics[width=0.5\textwidth]{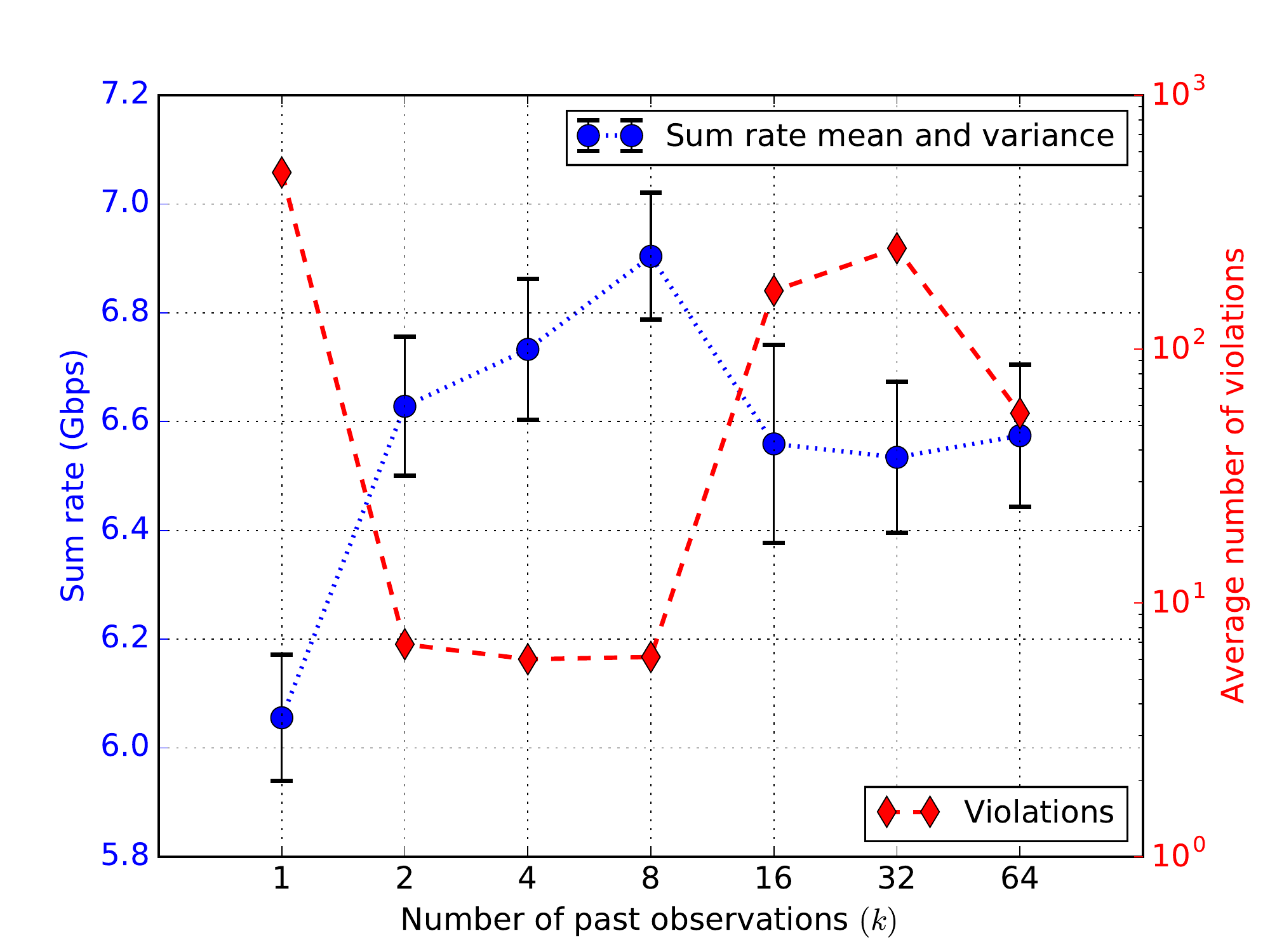}
	\caption{\label{history}Impact of utilizing channel history on sum rate and violations.}
\end{figure}

\subsection{{Neural Network Training}}
In this subsection, we analyze the impact of NN parameters on the performance of the proposed solution. The effect of history on the sum rate and average violations can be seen from Fig. \ref{history}. The history of channels plays an important role in the learning of vehicular mobility i.e., utilizing more history leads to more awareness of user mobility resulting into a mobility aware vehicle-RSU association policy. However, due to variation in mmWave channels, utilizing large number of past observations degrades the performance of the system. This is evident from the Fig. \ref{history}, where the utilization of eight past observation provide the maximum sum rate and minimum violations. On the other hand, when we increase the number of past observations to $16$, we see a decrease in the sum rate and an increase in the number of violations. The NN learns the correlation between the past observations and optimize its policy based on the observation. Since, we are utilizing mmWave links and finding the correlation for large number of channel observations require more training compared to the case with less channel realizations. 

\begin{figure}[hbtp]
	\centering
	\includegraphics[width=0.5\textwidth]{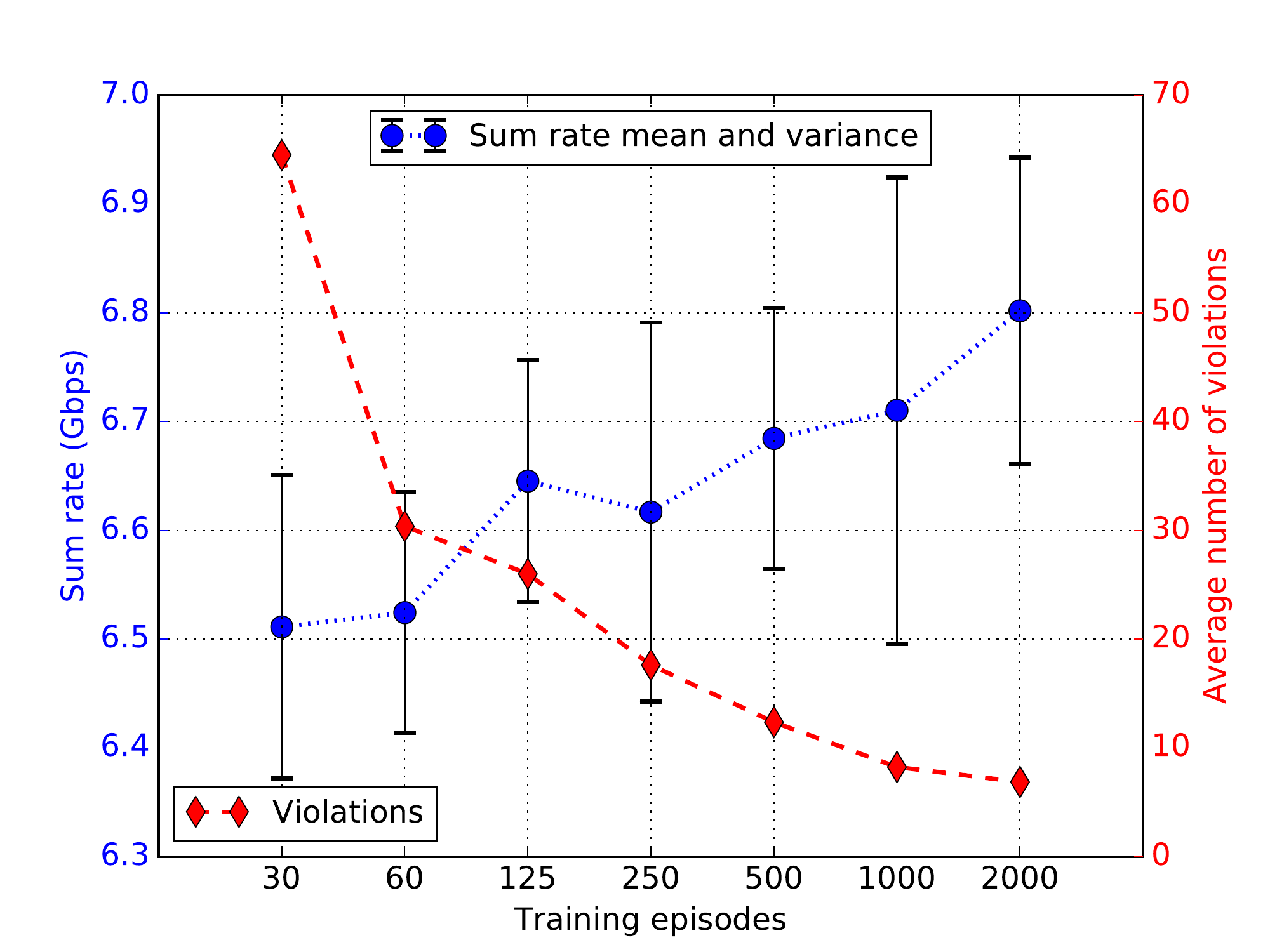}
	\caption{\label{episodes}Effect of training episodes on sum rate.}
\end{figure}

The NN in this work is trained in episodes, where one episode represent the sequence of agent interactions with the environment until a terminal state is reached. The terminal state here refers to the state when the vehicles leave the service area. The impact of training episodes is shown in Fig. \ref{episodes} and the performance is compared in terms of the sum rate and the average violations. The training episodes play an important role in converging to an optimal policy. We can see from Fig. \ref{episodes},  that lower value of training episodes result in minimum sum rate and the maximum number of violations. The number of violations decrease as we train the DRL agent i.e., the average number of violations for $30$ training episodes is $64$ and the number of violations reduces to $8$, when the training is performed for $2000$ episodes. Moreover, the sum rate also increases with the increase in training episodes.

\begin{figure}[hbtp]
	\centering
	\includegraphics[width=0.5\textwidth]{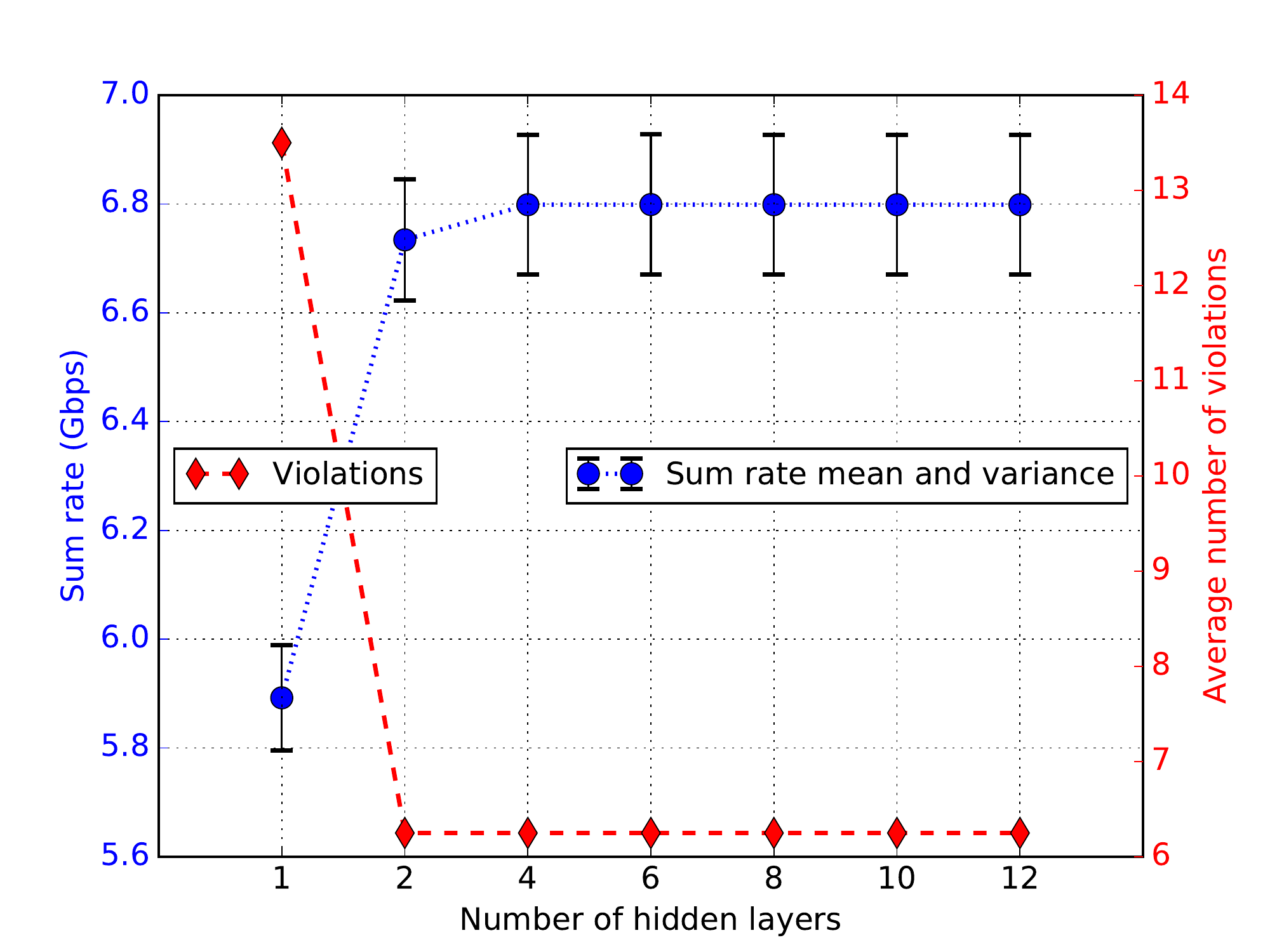}
	\caption{\label{layers}Impact of the number of NN hidden layers on sum rate and violations.}
\end{figure}

The number of hidden layers in the NN plays an important role. The impact of hidden layer on the proposed objective function is shown in Fig. \ref{layers}. The number of hidden layers can improve the accuracy depending upon the problem. However, increasing the number of hidden layers larger than the sufficient number of layers will cause accuracy in the test set to decrease. We can see from the Fig. \ref{layers}, that the average number of violations decrease from $13.5$ to $6.2$ when we increase the layers from $1$ to $2$ and after that increasing the number of layers does not decreases the violations. Moreover, we see an increase in the sum rate when the number of hidden layers are increased from $1$ to $4$ and after that increasing the number of layers does not have any effect on the sum rate. The number of hidden layers determines the capacity to learn the underlying patterns and we can see that the problem at hand can be learned by utilizing $4$ hidden layers.

\section{Conclusion\label{sec:Conclusion}}

In this paper, we proposed a reinforcement learning based distributed solution for vehicular user association in a V2X scenario. 
%We considered a scenario of six RSUs and eight vehicles operating within a service area, where each RSU is independently serving at most one vehicle at any given time. 
The goal is to devise vehicle-RSU association policy to enhance mobile user experience in terms of maximizing the network-wide sum rate while guaranteeing a minimum level of service rate for all vehicles. 
The vehicle-RSU association policy is found using distributed deep reinforcement learning techniques by utilizing observations of the rewards from past decisions across a large number of channel traces. 
This allows RSUs to learn the association policy for different network states. 
Numerical results demonstrate performance enhancement of the proposed solution over several state-of-the-art baselines models.     

\section*{Acknowledgment}
This research was supported by the Kvantum institute strategic project SAFARI, CARMA, MISSION, NOOR, SMARTER, High5 project number 2192/31/2016 funded by Business Finland, Bittium, Keysight, Kyynel, MediaTek, Nokia, University of Oulu and the Academy of Finland 6Genesis Flagship project under grant 318927.

\bibliographystyle{IEEEtran}
\bibliography{FINAL_VERSION}

\end{document}